\documentclass[aps,prb,twocolumn,superscriptaddress,floatfix,nofootinbib]{revtex4-1}
\usepackage[utf8]{inputenc}
\usepackage[dvipsnames]{xcolor}
\usepackage{graphicx}
\usepackage{amsmath}
\usepackage{multirow}
\usepackage{comment}
\usepackage{url}
\usepackage{pifont}

\usepackage{colortbl}

\usepackage{filecontents}

\usepackage{xr}
\makeatletter
\newcommand*{\addFileDependency}[1]{
  \typeout{(#1)}
  \@addtofilelist{#1}
  \IfFileExists{#1}{}{\typeout{No file #1.}}
}
\makeatother

\newcommand*{\myexternaldocument}[1]{%
    \externaldocument{#1}%
    \addFileDependency{#1.tex}%
    \addFileDependency{#1.aux}%
}
\myexternaldocument{si}

\begin{document}

\title{Theoretical investigation of decoherence channels in athermal phonon sensors}
\author{Thomas F. Harrelson}
\affiliation{Materials Science Division, Lawrence Berkeley National Laboratory, Berkeley, CA 94720, USA}
\affiliation{Molecular Foundry, Lawrence Berkeley National Laboratory, Berkeley, CA 94720, USA}
\author{Ibrahim Hajar}
\affiliation{Molecular Foundry, Lawrence Berkeley National Laboratory, Berkeley, CA 94720, USA}
\affiliation{Department of Mathematics, University of California - San Diego, CA 92093}
\affiliation{The Blackett Laboratory, Imperial College London, Prince Consort Road, London SW7 2BW, U.K.}
\author{Omar A. Ashour}
\affiliation{Materials Science Division, Lawrence Berkeley National Laboratory, Berkeley, CA 94720, USA}
\affiliation{Molecular Foundry, Lawrence Berkeley National Laboratory, Berkeley, CA 94720, USA}
\affiliation{Department of Physics, University of California, Berkeley, California 94720, USA}
\author{Sinéad M. Griffin}
\affiliation{Materials Science Division, Lawrence Berkeley National Laboratory, Berkeley, CA 94720, USA}
\affiliation{Molecular Foundry, Lawrence Berkeley National Laboratory, Berkeley, CA 94720, USA}

\begin{abstract}
 The creation and evolution of nonequilibrium phonons is central in applications ranging from cosmological particle searches to decoherence processes in qubits. However,  the fundamental understanding of decoherence pathways for athermal phonon distributions in solid-state systems remains an open question.  Using first-principles calculations, we investigate the primary decay channels of athermal phonons in two technologically relevant semiconductors -- Si and GaAs. We quantify the contributions of anharmonic, isotopic, and interfacial scattering in these materials. From this, we construct a model to estimate the thermal power in a readout scheme as a function of time. We discuss the implication of our results on noise limitations in current phonon sensor designs and strategies for improving coherence in next-generation phonon sensors. 
 
\end{abstract}

\maketitle

\section{Introduction}
%


Athermal phonon distributions are generated through many types of interactions in condensed matter physics. However, accessing  information in these distributions is challenging despite the value this information could bring to efforts including searches for cosmological particles like dark matter~\cite{supercdms20}, and for neutrino-less double beta decay experiments~\cite{dolinski_annrev19}. Efficient sensing of athermal phonons is also valuable for other scattering events that initially generate other quasiparticles such as excitons, photons, or magnons that readily down convert to phonons~\cite{antonius_arxiv17, krauss_prb96, li_natcomm14, berk_natcomm19, streib_prb19}. However, phonon decay processes lead to losses that limit overall detection efficiency. Mitigating these losses is challenging, but research on quantum acoustodynamic circuits has shown that some phonons have coherence times long enough to be used as carriers of quantum information~\cite{oconnell_nature10, chu_science17, chu_nature18, moores_prl18}, which motivates the study of coherence properties of phonons for sensing. 

\begin{figure}
    \centering
    \includegraphics{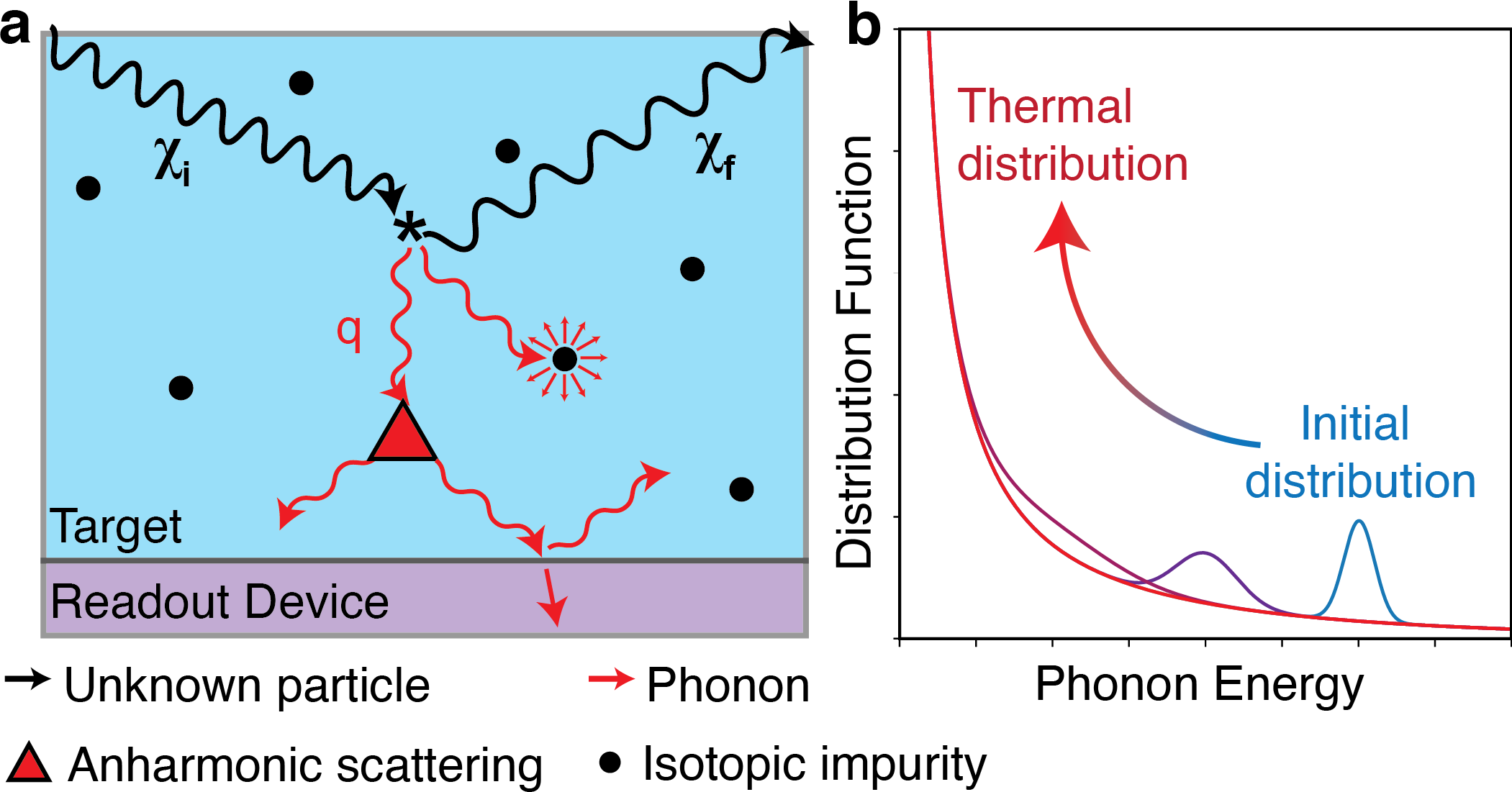}
    \caption{(a)Schematic of non-equilibrium phonon decay in a solid-state material  after the initial scattering event between an unknown particle ($\chi$) and the crystal. The initial population of phonons decays due to anharmonicities in the potential energy (red triangle), isotopic impurities (black dots), and interactions with the readout device surface. The readout device is usually an Al film attached to a transition edge sensor. (b) Time series of athermal phonon distribution in which the initial distribution loses average energy over time before eventually becoming indistinguishable from the thermal state.}
    \label{fig:sketch}
\end{figure}

Most non-equilibrium phonon sensors have a readout device attached to a phonon absorber material (also called a ``target'') that interacts with an external signal to generate non-equilibrium phonon distributions (see Figure~\ref{fig:sketch}). In searches for dark matter, these absorber materials are typically high-quality single-crystal semiconductors such as Si and Ge~\cite{cdmslite, supercdms20}. Increasing the design space for the types of target materials to lower dark matter masses is a recent focus of the community, with several alternative target materials already identified~\cite{knapen_prd17_sfhelium, griffin_21_sic, griffin_prd18, kurinsky_19_diamond, griffin_prd20, trickle_prl20, hochberg_prd18_dirac, Roising_et_al:2021, Inzani_et_al:2021}. Common readout devices include kinetic-inductance detectors (KIDs)~\cite{zmuidzinas_annrevcmp12}, metallic magnetic calorimeters~\cite{Fleischmann2005MetallicCalorimeters}, and transition-edge sensors (TES)~\cite{tes_review, pyle_apl18_tes}. In particular, TES devices are currently used in low-mass dark matter search experiments such as SuperCDMS and CRESST~\cite{supercdms20, cresst_ii, cresst_iii}, and KIDs have been used for cosmic microwave background experiments~\cite{zmuidzinas_nature}. Recent advances have improved the energy resolution of TES devices to energies comparable to that of a single optical phonon~\cite{fink_aip20}, which has critical implications for the detection of sub-MeV dark photons derived from GeV hidden sector dark matter~\cite{arkani_jhep08, cheung_prd09, morrissey_jhep09}. A dark photon is a particle that exists in a hypothetical sector of dark matter particles, and dark photons are kinetically mixed with ordinary photons~\cite{darkphoton}, which allow interactions with optical phonons near the Brillouin zone center. A current challenge for the detection of dark photons is that optical phonons near the Brillouin zone center have near-zero group velocities, and cannot be directly detected with current devices. As a result, the phonons created after optical phonon decay must be detected. 

The dominant phonon decay mechanisms in solid-state materials are due to anharmonic, isotopic, and surface interactions (see Figure~\ref{fig:sketch}). \textit{Ab initio} calculations, such as density functional theory (DFT)~\cite{HohenbergKohn1964, KohnSham1965}, are used to  calculate anharmonic and isotopic phonon scattering rates, accurately predicting properties such as heat transfer coefficients and electron/phonon lifetimes~\cite{phono3py, Tamura1983IsotopeGe, lindsay_prb14, stokes_prb11}. The key requirement for obtaining these properties is accurate calculations of the interatomic potential energy surfaces of the system  (e.g. three-phonon matrix elements) -- this can either be calculated perturbatively from static DFT calculations, or by sampling the atomic dynamics at a finite temperature using \textit{ab-initio} molecular dynamics (MD)\cite{ganose_natcomm_21, zhou_prl_14}. The latter observes the phonon dynamics directly at all orders of perturbation theory by calculating the dynamics of a large system over long enough time scales to incorporate the relevant phonons. This makes such MD methods more suitable for higher temperatures where typical phonon wavelengths and lifetimes are smaller than at low temperatures. For low-temperature phenomena, however, the construction of anharmonic matrix elements from static DFT calculations is preferable because quantum effects become more important at temperatures smaller than phonon frequencies. Previous work has found that while anharmonic interactions  impact  phonon dynamics at low temperatures ($<10$~K), the dominant contribution to heat transport at these cryogenic temperatures is surface scattering~\cite{glassbrenner_64, carlson_65}. However, a fully atomistic DFT calculation of the interface scattering with both chemical and structural specificity is computationally demanding owing to the large scale of the structures required. As a result, phenomenological models have been developed to describe surface scattering that incorporate materials-specific properties such as the phonon group velocity~\cite{morelli_prb02}, and interfacial properties such as the surface roughness~\cite{malhotra_scireps16}.

In this work, we use \textit{ab-initio} calculations to understand the role of different decay processes on athermal phonon distributions in two of the most prevalent  semiconductors in quantum technologies -- Si and GaAs~\cite{sipahigil_prl_14, sipahigil_prl_17, bluhm_natphys_11, cao_prl_16}. Si and GaAs are  current or near-term targets in phonon-based dark matter direct detection experiments, so understanding the phonon dynamics for these materials is especially timely. GaAs -- a polar semiconductor -- is of particular interest for dark matter detection since its optical phonons can couple to dark photons, providing a phonon-based pathway for probing dark-photon-based models of the dark sector\cite{griffin_prd18}. We first use DFT to calculate nonequilibrium phonon decoherence channels due to anharmonic, isotopic, and surface scattering. We next combine these results with a thermal transport model to calculate the athermal distribution as a function of time, and the rate of phonon energy transport across a readout device interface. We select a specific case that is especially relevant for dark matter detection, namely  the creation of a single optical phonon by dark-photon-mediated interactions (see  Figure~\ref{fig:sketch}b). This case is particularly important for exploring dark photon models in the terahertz range since they are predicted to generate a single optical phonon near the $\Gamma$ point of the Brillouin zone~\cite{griffin_prd20}. Finally, we discuss the implications of our materials-specific thermal transport model on single phonon detection and suggest routes to improving phonon coherence in such quantum sensing applications in near-term experiments. 

\section{Methods}

DFT calculations were performed with the Vienna Ab initio Simulation Package (VASP)~\cite{kresse93, kresse96a} using the PBE functional~\cite{PBE}. We chose the VASP recommended pseudopotentials in which the 4s, 4p electrons are treated as valence in Ga and As, and the 3s, 3p electrons are treated as valence in Si and Al; all other electrons were frozen into the core of the pseudopotential. We used a 600~eV plane-wave basis, and a $4\times4\times4$ Monkhorst-Pack grid of k-points, shifted from the $\Gamma$ point by half a grid point, for the conventional unit cell. Born effective charges were calculated within VASP using density functional perturbation theory~\cite{baroni01}. We used an electronic convergence criterion of $10^{-8}$~eV, and force minimization criterion of $2\times 10^{-4}$~eV/$\text{\AA}$. Phonon band structures were calculated with Phonopy~\cite{togo15} with the force constant matrices calculated with VASP using a $2\times2\times2$ supercell of the conventional standard cell, a k-point grid of $4\times4\times4$, and a q-point grid of $23\times23\times23$ for all three materials. Anharmonic scattering rates and interaction parameters (square modulus of anharmonic matrix elements) were calculated with Phono3py~\cite{phono3py}. 

Calculations of the phonon frequencies with isotopic variation were performed using isotopic supercells generated from our own code available at \url{https://github.com/IbraHajar/GaAs_Phonon_Frequencies}. In this code, we explicitly include isotopic variation by constructing supercells of atoms with distributions consistent with their natural abundances. The workflow for generating these explicit isotopic supercells is in the Appendix.  We wrote a code to build the thermal transport model which is available at \url{https://www.github.com/tfharrelson/MPDSF/scripts/readout_temperature_model}. All phonon-related parameters such as those used to construct the $\Gamma$ matrices in Eq.~\eqref{eq:g_partial} were calculated using either Phonopy or Phono3py. Numerical integration of Eq.~\eqref{eq:g_partial} was done via both forward and backward Euler's method. Forward Euler was used for the small time plots in Figure~\ref{fig:decay_model}a, and backward Euler was used to compute the curve in Figure~\ref{fig:decay_model}b. For both methods, we integrated $10^5$ time steps, and the time step was 1~ps, and $10^4$~ps for forward and backward Euler methods, respectively. We chose  a time step of 1~ps for the forward Euler method since it is a factor of 10 smaller than the lifetime of any optical phonon (the smallest time scale in the calculation). A time step of 10$^4$~ps was chosen for the backward Euler method since 10$^5$ time steps could be integrated in a reasonable timeframe to calculate transport up to 1~ms from the initial scattering event. As the backward Euler method is always stable with respect to the time step size, it is determined by computational requirements.

\section{Results}

\begin{figure*}
    \centering
    \includegraphics[width=\linewidth]{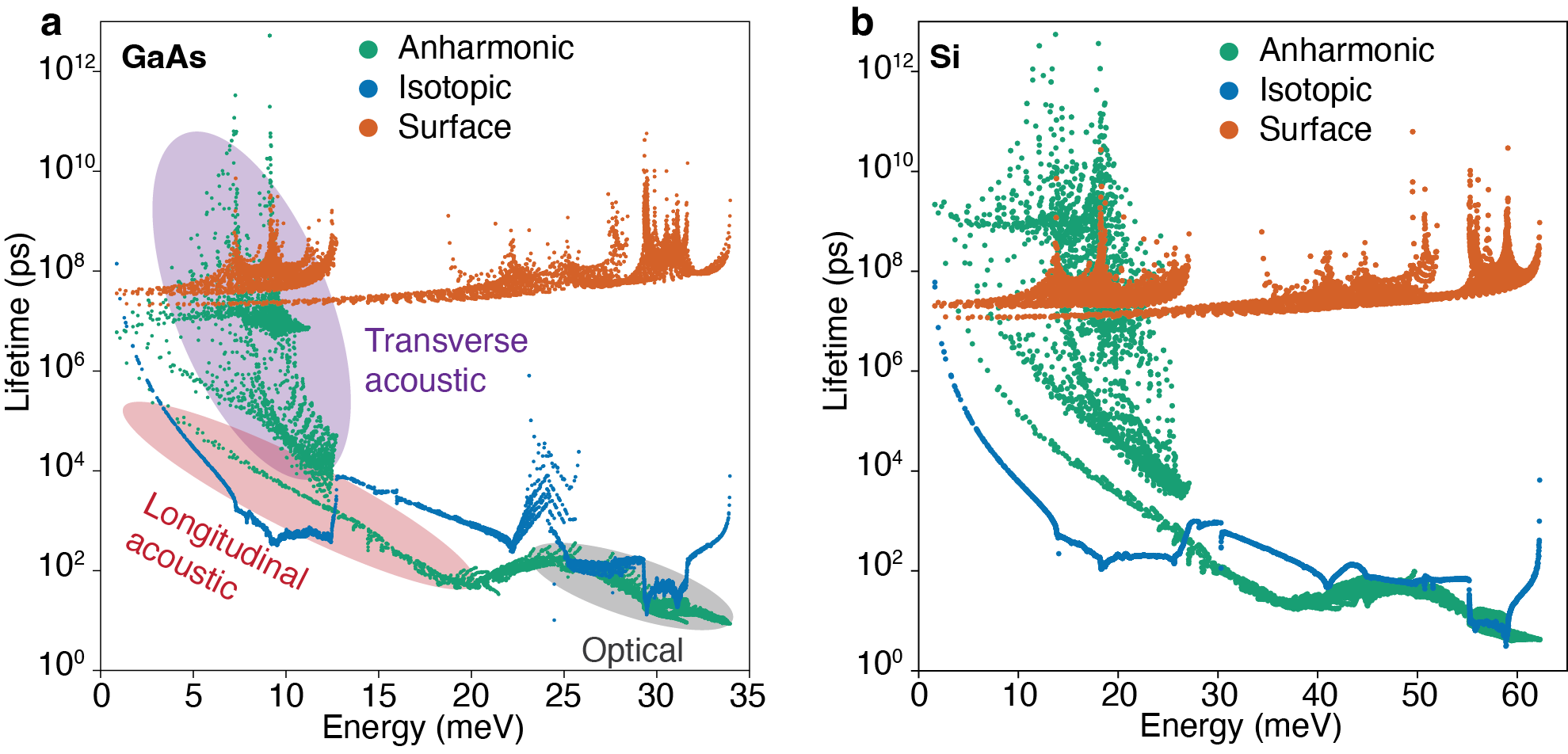}
    \caption{Calculated relaxation times for three different scattering processes (isotopic, anharmonic, and surface) for (a) GaAs, and (b) Si. GaAs anharmonic lifetimes associated with longitudinal acoustic, transverse acoustic, and optical modes are shaded purple, red, and grey, respectively. }
    \label{fig:scattering}
\end{figure*}

\subsection{First-principles calculations}
Si and GaAs adopt the diamond and zincblende crystal structures, respectively (see Supplementary Information). We calculated the lattice parameters to be 5.47~$\text{\AA}$ and 5.76~\AA, consistent with previous calculations~\cite{materialsproject} and experiments~\cite{encyclopedia} which have measured the lattice constants to be 5.43~$\text{\AA}$ and 5.65~\AA, respectively. The corresponding phonon band structures were also calculated (see SI, Fig. S1) and agree with previous phonon calculations using first-principles methods~\cite{materialsproject}. 

We next calculated the relaxation times associated with isotopic impurities~\cite{Tamura1983IsotopeGe} and anharmonic interactions~\cite{maradudin62} for GaAs and Si, which are shown in Figure~\ref{fig:scattering}. The isotopic scattering rates were calculated assuming the natural isotopic distributions of the relevant atoms (92.2\% $^{28}$Si, 4.7\% $^{29}$Si, 3.1\% $^{30}$Si, and 60.1\% $^{69}$Ga, 39.9\% $^{71}$Ga), and assuming that the positions of the isotopic impurities are uncorrelated. As expected, we find that the shapes of the distribution of the lifetimes are very similar between GaAs and Si owing to their similar crystal structure and phonon structure (see SI section \ref{si:sec:pbands} for more details). We find that the optical modes (shaded grey in Figure~\ref{fig:scattering}a) have relatively short anharmonic decay lifetimes of $10^1-10^2$~ps, which agrees with previous experiments~\cite{canonico_prl_02}, and suggest that the anharmonic interactions are the dominant decay process for optical modes in these materials. As the phonon energy decreases into the acoustic regions (shaded red and purple), we find that the lifetimes increase, including extremely large anharmonic lifetimes ($>1$~s) for transverse acoustic modes at $7-9$~meV for GaAs (shaded purple), and $15-20$~meV for Si. Since transverse acoustic modes typically have a finite group velocity, these modes propagate coherently over macroscopic distances in the absence of other decay processes. Therefore, these acoustic modes are the dominant information carriers of the original scattering event.

\subsection{Anharmonic Scattering of a Single Optical Phonon}

\begin{figure}[t]
    \centering
    \includegraphics[width=\linewidth]{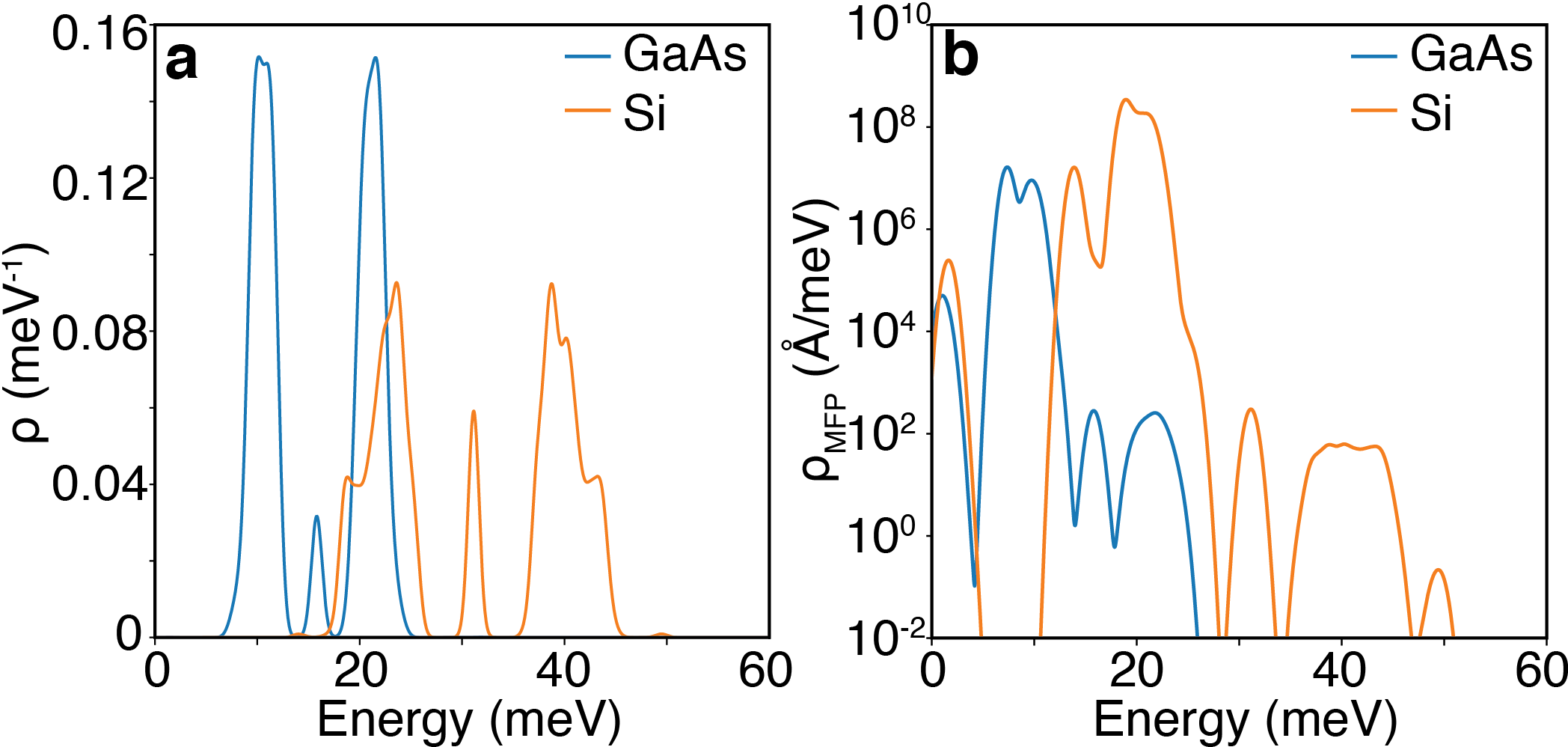}
    \caption{(a) The calculated anharmonic decay rate probability distribution ($\rho$) of the $\Gamma$-point optical mode for GaAs and Si.  (b) The probability distribution of the mean free path ($\rho_{MFP}$) of phonons that have been anharmonically scattered from the $\Gamma$-point optical modes for both GaAs and Si.}
    \label{fig:2nd-scatt}
\end{figure}

We next examine the scattering of a single optical phonon mode at the $\Gamma$ point from anharmonic scattering processes. As mentioned previously, this process is particularly relevant for dark photon models, and so we examine it more closely here. We first calculate the rate at which that optical phonon is anharmonically converted into lower energy phonons in Figure~\ref{fig:2nd-scatt}a, finding the optical phonon lifetime to be 10.9~ps from anharmonic contributions. We  calculate the optical phonon decay channels (interpreted as a probability density of a phonon decaying into a particular secondary channel) via,
\begin{widetext}
\begin{align} 
    \rho(\omega) =& \frac{r(\omega)}{\int r(\omega)\text{d}\omega} \label{eq:prob-density}
     \\
     \begin{split}
         r(\omega) =& \frac{18\pi}{\hbar^2} \sum_{\lambda_2,\lambda_3} |\phi_{\lambda_1, \lambda_2, \lambda_3}|^2\{(n_2 + n_3 + 1)\delta(\omega_1 - \omega_2 - \omega_3)\left[\delta(\omega - \omega_2) + \delta(\omega - \omega_3)\right] \\ 
    &+ (n_2 - n_3) [ \delta(\omega_1 + \omega_2 - \omega_3) (\delta(\omega - \omega_3) - \delta(\omega - \omega_2)) 
    - \delta(\omega_1 - \omega_2 + \omega_3) (\delta(\omega - \omega_2) - \delta(\omega - \omega_3)) ] \}
    \end{split}
\label{eq:ratebeast}
\end{align}
\end{widetext}
where $\phi_{\lambda_1,\lambda_2, \lambda_3}$ is the three-phonon anharmonic matrix element, $\lambda_i$ is the combined notation for both the branch index and k-point of a specific phonon mode, $n_i$ is the expectation value of the number of phonons in mode $\lambda_i$, $\omega_i$ is the frequency of mode $\lambda_i$, $r(\omega)$ is the frequency-dependent scattering rate, and $\rho(\omega)$ is called the decay rate probability distribution. In this equation, we group each of the phonons generated by the anharmonic matrix element into energy bins and sum over all allowed matrix elements before normalizing by the integral of the resulting function. In Figure~\ref{fig:2nd-scatt}a, we observe that the optical modes for GaAs decay most frequently into phonons at $\sim20$~meV and $\sim10$~meV and the optical modes for Si decay primarily into phonons at $\sim23$~meV and $\sim40$~meV. Inspection of the phonon band structures for both GaAs and Si (see SI Section \ref{si:sec:pbands}) concludes that all phonons generated by the anharmonic decay are acoustic phonons.

We next use the probability density defined in Eq.~\eqref{eq:prob-density} to construct  mean free path distributions (Figure~\ref{fig:2nd-scatt}b) for secondary scattered phonon populations. These are defined as,
\begin{equation}
    \rho_{MFP} = \frac{\lambda(\omega) r(\omega)}{\int r(\omega)\text{d}\omega}
\end{equation}
where $\lambda(\omega)$ is the average mean free path for each phonon with frequency~$\omega$. For clarity, we suppress the mode-specific relaxation times (mean free paths) in this expression, but to evaluate this they are inserted as weighting factors in the sum of Eq.~\eqref{eq:ratebeast} before normalizing against the integral of the rate distribution in Eq.~\eqref{eq:prob-density}. The large mean free paths in Figure~\ref{fig:2nd-scatt}b demonstrate that the low energy decay channel ($\sim$10~meV for GaAs and $\sim$23~meV for Si) includes a significant contribution of long-lifetime transverse acoustic modes in Figure~\ref{fig:scattering}, confirming that these modes are the dominant information carriers in the sensor in the absence of other scattering mechanisms. We also find that phonons centered at $\sim 1$~meV become relevant despite their apparent absence in Figure~\ref{fig:2nd-scatt}a since these are long-lived phonons compared to phonons in the other decay channels. 

\subsection{Isotopic Scattering Effects}

\begin{figure}
    \centering
    \includegraphics[width=\linewidth]{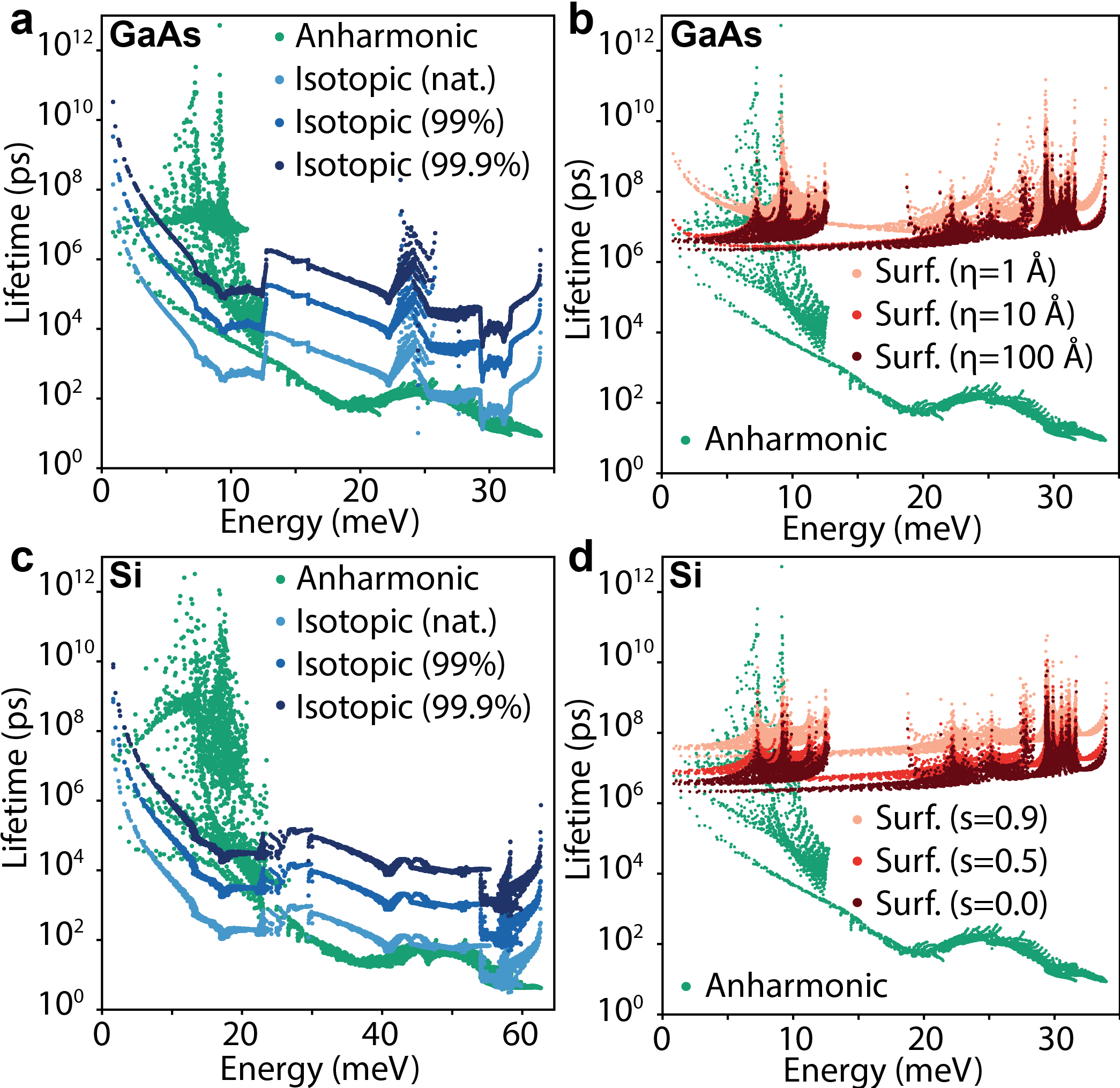}
    \caption{Calculated phonon lifetimes due to  isotopic scattering are plotted with increasing isotopic enrichment (light to dark blue dots) for GaAs (a) and Si (c). Calculated phonon lifetimes due to  surface scattering (dark to light red) are plotted with increasing values of roughness (b) and decreasing specularity (d) for GaAs. Calculated anharmonic lifetimes (green dots) are included in both plots for comparison.}
    \label{fig:surface_comp}
\end{figure}

As discussed above, while anharmonic contributions to phonon scattering are small for the low-energy decay channels of GaAs and Si, we find that the isotopic contributions  are relatively large (see Figure~\ref{fig:scattering}), and interfere with the low-energy decay channels of GaAs and Si. For both GaAs and Si, we find that the isotopic scattering is sub-dominant to anharmonic scattering for optical phonons, and becomes dominant for acoustic phonons. We next address the question of whether phonon scattering due to isotopic variation can cause phonons to thermalize (Figure~\ref{fig:sketch}b). At first glance, since isotopic scattering matrix elements couple two phonon fields, Fermi's golden rule would suggest that energy is conserved through the interaction, meaning that isotopic scattering cannot thermalize the phonon distribution. However, heat transport studies find that isotopic scattering rates must be included in a relaxation time model to accurately determine the temperature dependence of the phonon flux~\cite{lindsay_prb16}, suggesting that isotopic scattering rates cannot be ignored in thermalization processes. To explore this, we calculate the frequency of the $\Gamma$-point optical mode in GaAs for a series of random isotopic configurations of Ga masses (distributed according to the natural isotopic abundances; see SI section~\ref{si:sec:isotopes}). We find that the optical phonon frequency distribution has a finite width, implying that energy is not necessarily conserved in these isotopic scattering processes. In addition, this finite width is similar to  the isotopic scattering rate calculated from first principles. Since the isotopic decay rates lead to mean free paths that are much smaller than typical device sizes, we expect that the transport of phonons dominated by isotopic decay is diffusive, causing a substantial reduction in predicted thermal flux (see SI Section~\ref{si:sec:diffusive}).

Since we find that isotopic contributions to phonon decoherence need careful consideration for the design of next-generation athermal phonon sensors, we next explore how these scattering sources can be reduced. Mitigation of isotopic decay processes can be achieved through isotopic enrichment of the constituent elements prior to crystal growth. In the past, Si and GaAs have been enriched to 99.99\% and 99.4\% purity, respectively~\cite{veldhorst_natnano14, gaas_enrichment}. In Figures~\ref{fig:surface_comp}a and \ref{fig:surface_comp}c, we compare the calculated phonon decay lifetimes due to isotopic scattering at the natural isotopic abundance, at 99\% isotopic purity, and at 99.9\% isotopic purity. We find that the isotopic decay lifetimes become larger than the longitudinal acoustic lifetimes limited by anharmonic processes at 99\% isotopic purity for GaAs and 99.9\% for Si. At 99.9\% purity, the isotopic lifetimes become comparable to many of the transverse acoustic lifetimes for both materials, indicating that isotopic decay is no longer dominant -- but still relevant -- for those phonon modes.

\subsection{Surface Scattering Effects}

The final phonon decay processes we consider are those associated with surfaces and interfaces. We calculated the surface scattering rates using two different analytic models -- (1) a phenomenological model from ballistic phonon-interface interactions that account for all inelastic interactions (including defects) in a crude way, and (2) the Kirchhoff model that accounts for the impact of surface roughness on diffusive scattering while assuming there are no inelastic collisions from defects at the interface. The first follows a phenomenological equation,
\begin{equation} \label{eq:surf_spec}
    \frac{1}{\tau_{surf}} = \frac{v_g}{L}(1 - s)
\end{equation}
in which $\tau_{surf}$ is the surface scattering lifetime, $v_g$ is the group velocity of the phonon mode, $L$ is the length of the crystal, and $s$ is the specularity parameter determining the fraction of phonons that are specularly (elastically) reflected from the interface. In a previous study~\cite{morelli_prb02}, $s$ are not treated as independent variables and rather combined in a single $L$ fitting parameter. There, $L$ no longer represents the length and is equivalent to fixing $L$ and varying $s$ in our Eq.~\eqref{eq:surf_spec}. In Figure~\ref{fig:scattering}, we compare the surface scattering lifetimes with a specularity parameter of 0.9 and a 1~cm crystal size against the anharmonic and isotopic lifetimes. A specularity parameter of 0.9 is equivalent to asserting that the phonon mean free path is 10-fold greater than the crystal size, which reasonably agrees with experimentally measured mean free paths of phononic metamaterials~\cite{alaie_natcomm15}. We find that the surface scattering rates are typically around 0.1~ms, which is greater than all other lifetimes considered in both materials, aside from the highest transverse acoustic anharmonic lifetimes. In this case, the non-specular surface scattering rates are the limiting factor in the overall phonon lifetime.

One route to reduce surface scattering is to fabricate exceptionally clean and flat surfaces (hence reducing $s$). To gain more insight into the effect of surface roughness on the scattering specularity, we consider the Kirchhoff surface model~\cite{malhotra_scireps16},
\begin{equation} \label{eq:kirchoff}
    s_k = \frac{\pi}{4\eta k} \text{erf}\left(2\eta k\right)
\end{equation}
which relates $s_{k}$ (the momentum-dependent specularity parameter) to $\eta$, the surface roughness (with units of length), and $k$,  the phonon wavevector incident to the interface. Derivation of this equation is given in the SI Section~\ref{si:sec:kirchhoff}. The scattering lifetimes are computed by replacing $s$ in Eq.~\eqref{eq:surf_spec} with $s_k$ above. In this model, the scattering interactions decrease as the phonon wavelength exceeds the length scale of the interface deformation.

In Figure~\ref{fig:surface_comp},  we plot our calculated lifetimes for phonons in GaAs with different values of surface roughness, $\eta$ (Figure~\ref{fig:surface_comp}b), and a constant specularity parameter, $s$ (Figure~\ref{fig:surface_comp}d). We find that a roughness of 1 \AA~generates lifetimes similar to that of a constant specularity parameter of 0.9 in Figure~\ref{fig:surface_comp}d. While the magnitudes of the lifetimes are similar, we find that the low-energy shapes are qualitatively different -- the surface lifetimes Eq.~\eqref{eq:kirchoff} increase with decreasing energy, while the lifetimes from Eq.~\eqref{eq:surf_spec} remain constant with decreasing energy. This difference is due to the strength of the scattering interaction from the Kirchhoff model scaling with the phonon wavevector. 

In Figure~\ref{fig:surface_comp}b, we show that the onset of this scaling behavior increases in energy as  the surface roughness ($\eta$) decreases. For a surface roughness of $\eta=100$~\AA, the lifetimes become very similar to the $s=0$ lifetimes in Figure~\ref{fig:surface_comp}d, indicating that the phenomenological model in Eq.~\eqref{eq:surf_spec} and the Kirchhoff model coincide for rough interfaces. We expect that a combination of both models is required to explain the surface scattering in experiments because all fabricated material interfaces have some roughness and some defects leading to inelastic scattering. For example, assuming a pure Kirchhoff model could lead to a very inaccurate characterization of the surface scattering if defect scattering rates dominate. The exact details of the surface roughness, defect profile, etc. of the fabricated system are therefore needed for more accurate predictions. 

Despite the lack of surface information for our materials, our results in Figure~\ref{fig:surface_comp} have enough information to draw reasonable conclusions regarding the critical scattering rates in GaAs and Si. For the $s=0.9$ and $\eta=1$~\AA\ cases of the phenomenological and Kirchhoff models, respectively, the phonon lifetimes due to surface scattering are between $10^7-10^9$~ps, which is significantly larger than many of the anharmonic lifetimes. This indicates that surface scattering is not expected to dominate with reasonably well-polished surfaces until the athermal phonon distribution down converts below $\sim1$~meV. Experiments have shown that Si surface roughnesses are of the order of $0.1-1$~nm~\cite{teichert_apl95}, suggesting that the surface scattering rate is small for well-studied materials like Si and GaAs.


\subsection{Thermal transport model}

\begin{figure}
    \centering
    \includegraphics[width=\linewidth]{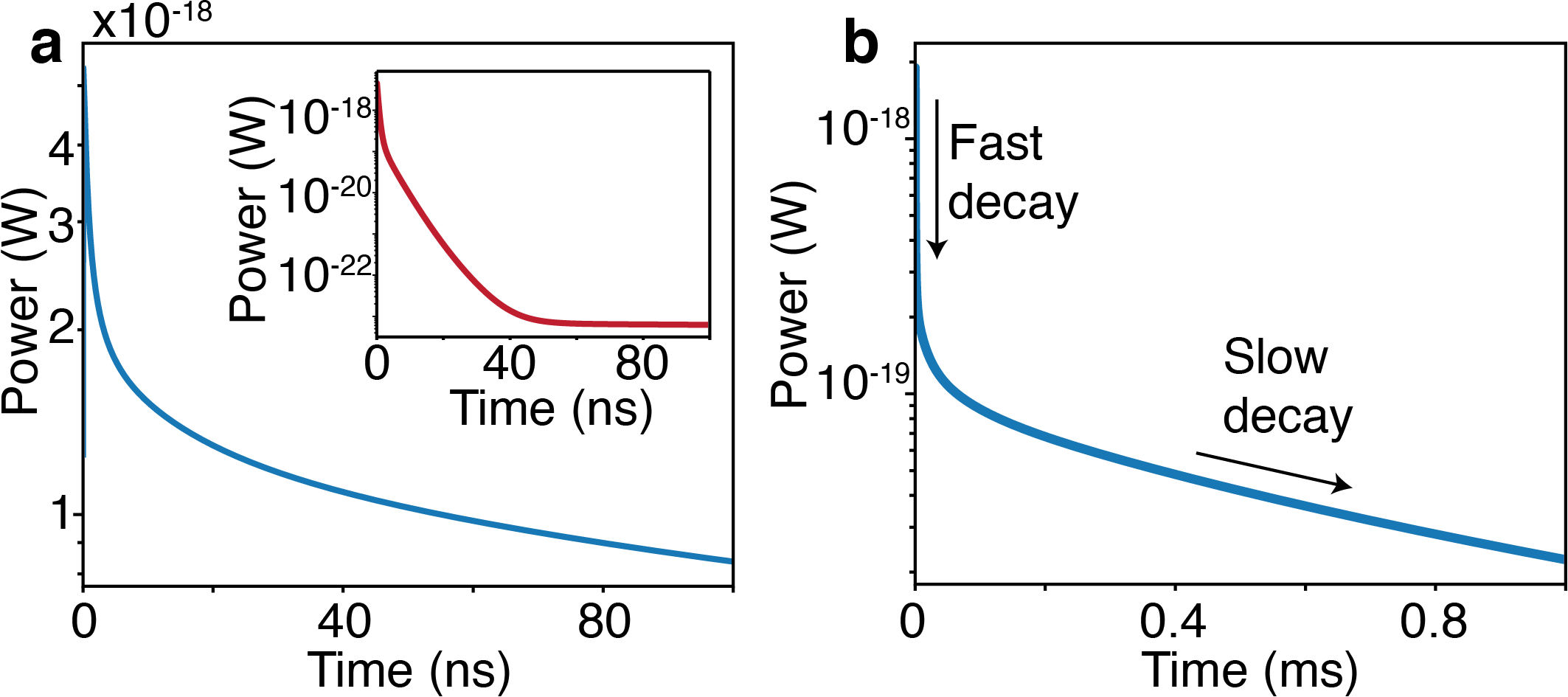}
    \caption{(a) Thermal power transmitted across GaAs/readout device interface as a function of time without the impacts of isotopic scattering. Inset is the transmitted thermal power including the impacts of isotopic scattering. (b) The long-time behavior of the transmitted thermal power as a function of time with three distinct scaling regions identified. }
    \label{fig:decay_model}
\end{figure}
We next construct a model to describe the energy transport into a readout device originating from a non-equilibrium $\Gamma$-point optical phonon distribution in a macroscopically-sized GaAs crystal
using the \textit{ab initio} parameters. The equation for the heat flux~\cite{swartz_pohl_revmodphys_89} is,
\begin{equation}
    \phi(t) = \frac{1}{2}\int \hbar\omega \langle v_g(\omega)\rangle D(\omega) \alpha(\mu, \omega) \mu g(\omega, t) \text{d} \mu \text{d}\omega
\end{equation}
where $\langle v_g(\omega)\rangle$ is the mean group velocities of modes at $\omega$, $D(\omega)$ is the phonon density of states, $\alpha(\mu, \omega)$ is the transmission coefficient of phonons, $g(\omega, t)$ is the non-equilibrium part of the full phonon distribution, and $\mu=\cos\theta$, where $\theta$ is the angle between the phonon wavevector and the surface normal. We first set $\alpha=0.5$, that is, we assume an average value of $\alpha$, which is reasonable since $\alpha$ is the only $\mu$-dependent quantity in the integrand. The value of 0.5 is consistent with another study~\cite{harrelson_interface}. Averaging over $\mu$ yields another factor of 0.5, which simplifies the previous equation to,
\begin{equation} \label{eq:flux_model}
    \phi(t) = \frac{1}{8}\int \hbar\omega \langle v_g(\omega)\rangle D(\omega) g(\omega, t) \text{d}\omega.
\end{equation}
Normalization of the crystal volume sets the correct scale with  $\int D(\omega) g(\omega, 0)\text{d}\omega = 1/V$ in the limit that a single phonon is created in the crystal.  We assume a cubic crystal volume of 50~cm$^3$ (similar to crystal absorbers in the CRESST dark matter experiment~\cite{cresst_iii}), and a crystal surface coverage of 2.7\% for the readout (consistent with transmission edge sensor coverage~\cite{fink_aip20}). 

To calculate the time-dependent thermal power, $\phi(t)\times A$, ($A$ = surface area), we evolve $g(\omega, t)$ with time according to the phonon scattering rates in the material, via,
\begin{equation} \label{eq:g_partial}
    \frac{\text{d}g(\omega, t)}{\text{d} t} = -\Gamma_{anh}(\omega, \omega')g(\omega', t) - \Gamma_{iso}(\omega) g(\omega, t)
\end{equation}
where $\Gamma_{anh}(\omega, \omega')$ is the rate at which $g(\omega')$ is converted to $g(\omega)$ due to anharmonic scattering, and $\Gamma_{iso}(\omega)$ is the rate of isotopic decoherence at frequency $\omega$. We break $\Gamma_{anh}$ into diagonal and off-diagonal components, which represent the decay from an initial state (diagonal components) and the transfer to other states at different energies (off-diagonal components). The diagonal components are approximated as,
\begin{equation}
    \Gamma_{anh}(\omega,\omega) = \langle \tau_{anh}^{-1} (\omega)\rangle
\end{equation}
where $\langle \tau_{anh}^{-1} (\omega)\rangle$ is the average inverse anharmonic lifetime of the phonons created at frequency $\omega$ in Figure~\ref{fig:2nd-scatt}. Similarly, we assume the isotopic scattering rates to be purely diagonal, which we find by replacing the anharmonic lifetimes with isotopic lifetimes in the equation above. The off-diagonal terms in $\Gamma_{anh}$ represent the growth of the athermal distribution at $\omega$ at the expense of the decay at $\omega'$, and are provided by the scattering rates in Figure~\ref{fig:2nd-scatt}a. This implicitly assumes that rearrangement of the non-equilibrium distribution function only occurs once from the optical $\Gamma$-point mode into the decay channels in Figure~\ref{fig:2nd-scatt}a, and that any subsequent scattering, either anharmonic or isotopic, directly leads to thermalization.

Having defined all of the terms in Eq.~\eqref{eq:g_partial}, we numerically integrate $g(\omega, t)$, and insert the time-dependent values into Eq.~\eqref{eq:flux_model} to find the thermal power that crosses the readout interface as a function of time. In Figure~\ref{fig:decay_model}a, we plot the short-time results for the thermal power excluding the impact of isotopic decoherence. The initial power is $5\times 10^{-18}$~W, before quickly decaying to half the initial value over a time scale on the order of $\sim$1~ns, and slowly decaying further over 100s of ns. The scale of the predicted thermal power is similar to $1.5\times 10^{-18}$~W/Hz$^{1/2}$, which is the reported noise equivalent power of a current TES device~\cite{fink_aip20}. Adding isotopic decoherence to the thermal power model (inset Figure~\ref{fig:decay_model}a) removes the slow decay process, and the predicted thermal power decreases to $\sim10^{-23}$~W after 40~ns. 

We observe that the isotopic-free thermal power does not completely decay after 100~ns, suggesting that there are transport channels that are active for time scales longer than 100~ns. To examine this, we show the long-time scaling behavior of the power in Figure~\ref{fig:decay_model}b. We find two distinct scaling regions separated at $\sim1$~ns, which represent the two peaks in Figure~\ref{fig:2nd-scatt}b near 20~meV (fast decay), and 10~meV (slow decay), respectively. The linear relationship in Figure~\ref{fig:decay_model}b shows that the exponential decay process has an effective lifetime of 0.77~ms, which is comparable to the timescale associated with the 2.6~kHz bandwidth for a TES readout device.

\section{Discussion}
We find that isotopic scattering in non-enriched GaAs and Si-based sensors will severely diminish their performance. The reason for this is illustrated in Figure~\ref{fig:scattering}; the isotopic lifetimes are significantly smaller than anharmonic lifetimes for the optical phonon decay channels with large mean free paths near 10~meV and 20~meV in GaAs and Si, respectively. The impact of isotopic decay is clearly observed when we include isotopic interactions in the thermal transport model. The thermal power signal reaches peak power on the order of 10$^{-18}$~W, and the signal only survives for $\sim$1~ns. Given that the noise-equivalent power of a TES readout device is $1.5\times10^{-18}$~W/Hz$^{1/2}$ with a bandwidth of 2.6~kHz~\cite{fink_aip20}, the effective noise power is 7.6$\times 10^{-17}$~W, making the signal from a single optical phonon  undetectable without significant isotopic enrichment with current TES limits. For GaAs, assuming enrichment to $99.99\%$ purity, then the phonons near 10~meV have $\sim10^7$~ps lifetimes. Assuming that the characteristic velocity is $\sim 10^5$~cm/s, then the mean free path becomes $\sim 1$~cm. Therefore, we find that $\sim99.99\%$ purity is necessary to successfully mitigate the isotopic decoherence for $\sim1$~cm size GaAs absorbers. We continue our analysis of the feasibility of single phonon detection assuming the isotopic impurity-free case (or sufficiently enriched).

We find that, even in the limit of zero isotopic scattering, the predicted thermal power is too weak to be observed in current TES devices. Specifically, a single optical phonon in a 50~cm$^3$ absorber creates a thermal power impulse starting near $5\times10^{-18}$~W in the absence of isotopic scattering. The initial signal decays quickly over $\sim5$~ns, which is not long enough to be detected with a TES readout device that has a noise equivalent power of $1.5\times10^{-18}$~W/Hz$^{1/2}$ operating with a bandwidth of 2.6~kHz. Additionally, the noise power with this bandwidth is greater than the initial power from the phonons. The slow decay process allows the predicted signal to survive for 100s of $\mu$s, suggesting  a 1~MHz bandwidth should ideally be used. However, the corresponding power is between 10$^{-20}-10^{-19}$~W, which is 2-3 orders of magnitude smaller than the effective noise power of the readout device. 

Despite the substantial readout noise relative to the predicted signal, we note that there are several assumptions in the thermal transport model that can impact our predictions. The first is due to the simplifications made to the integration of the non-equilibrium distribution function. We assume that one of two things happens after the initial optical phonon decays: (1) the phonon transports out of the absorber to the readout device, or (2) the phonon decays to the thermal background. There is an intermediate option in which the secondary phonon population anharmonically decays again into lower-energy acoustic phonons that contribute to the thermal power. We are also assuming that the calculated phonon lifetimes represent a reasonable approximation to the actual scattering rate in a real material. Prior work comparing phonon lifetimes observed from Raman and inelastic neutron scattering experiments with DFT calculations show that the agreement is very reasonable as long as the temperature is small compared to the phonon frequency~\cite{debernardi_prl95, glensk_prl19}. The temperature of the detector in a sensor experiment is $10-100$~mK (equivalent to 1-10~$\mu$eV), which is much smaller than the phonon energies that we calculate. Thus, we expect that our first-principles calculations of phonon lifetimes are accurate, and any errors inherent to the calculation procedure are subdominant to other sources of error. As a result, our results in Figure~\ref{fig:decay_model}a represent a reasonable first approximation that is a lower bound for the actual result. 

While phonons that are generated from a small number of decay events contribute to thermal flux, phonons that arise after many decay events are not expected to contribute to the thermal flux. This is because the phonon momentum relaxes over several scattering processes, causing diffusive transport (instead of ballistic transport for coherent phonons) out of the absorber which has significantly reduced transport rates (see SI Section \ref{si:sec:diffusive} for details). Diffusive transport can be avoided by decreasing the effective phonon travel distance to become comparable to the mean free paths in Figure~\ref{fig:2nd-scatt}.

Additionally, the volume of the absorber and surface coverage of the readout device have a critical impact on the observed thermal power in Figure~\ref{fig:decay_model}. Linearly decreasing the volume of the absorber, causes a linear increase in the observed thermal power because the surface area also decreases with the volume. Another consideration is that the surface scattering rate scales with $V^{-1/3}$, meaning that surface scattering effects become much more relevant as the absorber volume decreases. The volume can be held constant while increasing the surface area of the readout device by changing the form factor from a cube to a film. This has the benefit of decreasing one of the dimensions of the absorber to be closer to more of the phonon mean free paths, which increases the collection efficiency. However, as the surface area to volume ratio increases, the expected residence time of a phonon decreases, which decreases the allowed integration time to collect the observed signal. Therefore, decreasing the surface area to volume ratio increases the noise bandwidth, causing an increase in the effective noise power. Thus, optimization studies including such geometric effects are needed to determine the number and size of detectors required to achieve a targeted sensitivity threshold.

\section{Conclusions and Outlook}
We calculate the anharmonic, isotopic, and surface scattering rates for both GaAs and Si and find that anharmonic scattering dominates for optical phonons and high-energy acoustic phonons. Isotopic scattering begins to dominate at 12~meV and 25~meV for GaAs and Si, respectively. This type of scattering is significantly detrimental to the predicted thermal power crossing the readout device interface. This can be significantly mitigated with isotopic enrichment greater than 99.9\%. We also find that surface scattering is not a dominant decay channel for most phonons modes, particularly when surface roughnesses are near 1~\AA. 

In the limit of zero isotopic scattering, the thermal power from a single optical phonon in a 50~cm$^3$ device reaches power levels comparable to that of current TES devices. However, the time scale for the signal decay is too small to overcome the effective noise power over small time scales in TES devices. Further device optimization is required to increase the observed signal-to-noise ratio. One possibility for design is to reduce the detector volume, while simultaneously increasing the interfacial area between the readout device and the absorber to maximize the initial signal magnitude.

The preceding discussion is based on the limit of zero isotopic scattering in the absorber, which is mitigated through isotopic enrichment. Another path for mitigation of isotopic scattering noise is to choose a material with constituent atoms that have no natural isotopic variation. Of the possible materials, aluminum arsenide (AlAs) seems to be a viable candidate that is most similar to GaAs and Si. AlAs is a polar semiconductor, and forms into a crystal with the same space group (zincblende, F$\overline{4}3$m) as GaAs. Because of this, AlAs's sensitivity to dark matter models should be similar to that of GaAs\cite{griffin_prd20}, in addition to similar anharmonic scattering properties as GaAs and Si. Since Al is lighter than Ga, the phonon energies and group velocities of AlAs are larger (see SI, Fig. S1(b) and (c)). Since the thermal power is proportional to group velocity, this should correspondingly increase the thermal power relative to the noise. The transmission coefficient should also increase from the improved impedance mismatch between AlAs and Al.  Also, single crystals of AlAs have been fabricated and used in waveguide devices~\cite{gourley86}. Thus, AlAs could feasibly produce detectable signals for single optical phonons with additional device optimization and/or advances in readout technology. The final option for increasing the predicted thermal power is to search for another material with anomalously low anharmonic interactions.

\section{Acknowledgments}
We thank Sean Lubner, Aritoki Suzuki, Matt Pyle, Kathryn Zurek, and Maurice Garcia-Sciveres for useful discussions. This work was supported by the Quantum Information Science Enabled Discovery (QuantISED) for High Energy Physics (KA2401032). Work at the Molecular Foundry was supported by the Office of Science, Office of Basic Energy Sciences, of the U.S. Department of Energy under Contract No. DE-AC02-05CH11231.  This research used resources of the National Energy Research Scientific Computing Center (NERSC), a U.S. Department of Energy Office of Science User Facility operated under Contract No. DE-AC02-05CH11231. This work used the Extreme Science and Engineering Discovery Environment (XSEDE), which is supported by National Science Foundation grant number ACI-1548562. IH acknowledges support from the DOE Science Undergraduate Laboratory Internship (SULI) program.

\bibliography{ref2}

\begin{thebibliography}{68}%
\makeatletter
\providecommand \@ifxundefined [1]{%
 \@ifx{#1\undefined}
}%
\providecommand \@ifnum [1]{%
 \ifnum #1\expandafter \@firstoftwo
 \else \expandafter \@secondoftwo
 \fi
}%
\providecommand \@ifx [1]{%
 \ifx #1\expandafter \@firstoftwo
 \else \expandafter \@secondoftwo
 \fi
}%
\providecommand \natexlab [1]{#1}%
\providecommand \enquote  [1]{``#1''}%
\providecommand \bibnamefont  [1]{#1}%
\providecommand \bibfnamefont [1]{#1}%
\providecommand \citenamefont [1]{#1}%
\providecommand \href@noop [0]{\@secondoftwo}%
\providecommand \href [0]{\begingroup \@sanitize@url \@href}%
\providecommand \@href[1]{\@@startlink{#1}\@@href}%
\providecommand \@@href[1]{\endgroup#1\@@endlink}%
\providecommand \@sanitize@url [0]{\catcode `\\12\catcode `\$12\catcode
  `\&12\catcode `\#12\catcode `\^12\catcode `\_12\catcode `\%12\relax}%
\providecommand \@@startlink[1]{}%
\providecommand \@@endlink[0]{}%
\providecommand \url  [0]{\begingroup\@sanitize@url \@url }%
\providecommand \@url [1]{\endgroup\@href {#1}{\urlprefix }}%
\providecommand \urlprefix  [0]{URL }%
\providecommand \Eprint [0]{\href }%
\providecommand \doibase [0]{http://dx.doi.org/}%
\providecommand \selectlanguage [0]{\@gobble}%
\providecommand \bibinfo  [0]{\@secondoftwo}%
\providecommand \bibfield  [0]{\@secondoftwo}%
\providecommand \translation [1]{[#1]}%
\providecommand \BibitemOpen [0]{}%
\providecommand \bibitemStop [0]{}%
\providecommand \bibitemNoStop [0]{.\EOS\space}%
\providecommand \EOS [0]{\spacefactor3000\relax}%
\providecommand \BibitemShut  [1]{\csname bibitem#1\endcsname}%
\let\auto@bib@innerbib\@empty
\bibitem [{\citenamefont {Amaral}\ \emph {et~al.}(2020)\citenamefont {Amaral},
  \citenamefont {Aralis}, \citenamefont {Aramaki}, \citenamefont {Arnquist},
  \citenamefont {Azadbakht}, \citenamefont {Banik}, \citenamefont {Barker},
  \citenamefont {Bathurst}, \citenamefont {Bauer}, \citenamefont {Bezerra},\
  and\ \citenamefont {et~al.}}]{supercdms20}%
  \BibitemOpen
  \bibfield  {author} {\bibinfo {author} {\bibfnamefont {D.~W.}\ \bibnamefont
  {Amaral}}, \bibinfo {author} {\bibfnamefont {T.}~\bibnamefont {Aralis}},
  \bibinfo {author} {\bibfnamefont {T.}~\bibnamefont {Aramaki}}, \bibinfo
  {author} {\bibfnamefont {I.~J.}\ \bibnamefont {Arnquist}}, \bibinfo {author}
  {\bibfnamefont {E.}~\bibnamefont {Azadbakht}}, \bibinfo {author}
  {\bibfnamefont {S.}~\bibnamefont {Banik}}, \bibinfo {author} {\bibfnamefont
  {D.}~\bibnamefont {Barker}}, \bibinfo {author} {\bibfnamefont
  {C.}~\bibnamefont {Bathurst}}, \bibinfo {author} {\bibfnamefont {D.~A.}\
  \bibnamefont {Bauer}}, \bibinfo {author} {\bibfnamefont {L.~V.~S.}\
  \bibnamefont {Bezerra}}, \ and\ \bibinfo {author} {\bibnamefont {et~al.}},\
  }\href {\doibase 10.1103/physrevd.102.091101} {\bibfield  {journal} {\bibinfo
   {journal} {Physical Review D}\ }\textbf {\bibinfo {volume} {102}} (\bibinfo
  {year} {2020}),\ 10.1103/physrevd.102.091101}\BibitemShut {NoStop}%
\bibitem [{\citenamefont {Dolinski}\ \emph {et~al.}(2019)\citenamefont
  {Dolinski}, \citenamefont {Poon},\ and\ \citenamefont
  {Rodejohann}}]{dolinski_annrev19}%
  \BibitemOpen
  \bibfield  {author} {\bibinfo {author} {\bibfnamefont {M.~J.}\ \bibnamefont
  {Dolinski}}, \bibinfo {author} {\bibfnamefont {A.~W.}\ \bibnamefont {Poon}},
  \ and\ \bibinfo {author} {\bibfnamefont {W.}~\bibnamefont {Rodejohann}},\
  }\href {\doibase 10.1146/annurev-nucl-101918-023407} {\bibfield  {journal}
  {\bibinfo  {journal} {Annual Review of Nuclear and Particle Science}\
  }\textbf {\bibinfo {volume} {69}},\ \bibinfo {pages} {219} (\bibinfo {year}
  {2019})}\BibitemShut {NoStop}%
\bibitem [{\citenamefont {Antonius}\ and\ \citenamefont
  {Louie}(2017)}]{antonius_arxiv17}%
  \BibitemOpen
  \bibfield  {author} {\bibinfo {author} {\bibfnamefont {G.}~\bibnamefont
  {Antonius}}\ and\ \bibinfo {author} {\bibfnamefont {S.~G.}\ \bibnamefont
  {Louie}},\ }\href@noop {} {\enquote {\bibinfo {title} {Theory of the
  exciton-phonon coupling},}\ } (\bibinfo {year} {2017}),\ \Eprint
  {http://arxiv.org/abs/1705.04245} {arXiv:1705.04245 [cond-mat.mtrl-sci]}
  \BibitemShut {NoStop}%
\bibitem [{\citenamefont {Krauss}\ and\ \citenamefont
  {Wise}(1997)}]{krauss_prb96}%
  \BibitemOpen
  \bibfield  {author} {\bibinfo {author} {\bibfnamefont {T.~D.}\ \bibnamefont
  {Krauss}}\ and\ \bibinfo {author} {\bibfnamefont {F.~W.}\ \bibnamefont
  {Wise}},\ }\href {\doibase 10.1103/PhysRevB.55.9860} {\bibfield  {journal}
  {\bibinfo  {journal} {Phys. Rev. B.}\ }\textbf {\bibinfo {volume} {55}},\
  \bibinfo {pages} {9860} (\bibinfo {year} {1997})}\BibitemShut {NoStop}%
\bibitem [{\citenamefont {Li}\ \emph {et~al.}(2014)\citenamefont {Li},
  \citenamefont {Eggleton}, \citenamefont {Fang},\ and\ \citenamefont
  {Fan}}]{li_natcomm14}%
  \BibitemOpen
  \bibfield  {author} {\bibinfo {author} {\bibfnamefont {E.}~\bibnamefont
  {Li}}, \bibinfo {author} {\bibfnamefont {B.~J.}\ \bibnamefont {Eggleton}},
  \bibinfo {author} {\bibfnamefont {K.}~\bibnamefont {Fang}}, \ and\ \bibinfo
  {author} {\bibfnamefont {S.}~\bibnamefont {Fan}},\ }\href {\doibase
  10.1038/ncomms4225} {\bibfield  {journal} {\bibinfo  {journal} {Nature
  Communications}\ }\textbf {\bibinfo {volume} {5}},\ \bibinfo {pages} {3225}
  (\bibinfo {year} {2014})}\BibitemShut {NoStop}%
\bibitem [{\citenamefont {Berk}\ \emph {et~al.}(2019)\citenamefont {Berk},
  \citenamefont {Jaris}, \citenamefont {Yang}, \citenamefont {Dhuey},
  \citenamefont {Cabrini},\ and\ \citenamefont {Schmidt}}]{berk_natcomm19}%
  \BibitemOpen
  \bibfield  {author} {\bibinfo {author} {\bibfnamefont {C.}~\bibnamefont
  {Berk}}, \bibinfo {author} {\bibfnamefont {M.}~\bibnamefont {Jaris}},
  \bibinfo {author} {\bibfnamefont {W.}~\bibnamefont {Yang}}, \bibinfo {author}
  {\bibfnamefont {S.}~\bibnamefont {Dhuey}}, \bibinfo {author} {\bibfnamefont
  {S.}~\bibnamefont {Cabrini}}, \ and\ \bibinfo {author} {\bibfnamefont
  {H.}~\bibnamefont {Schmidt}},\ }\href {\doibase 10.1038/s41467-019-10545-x}
  {\bibfield  {journal} {\bibinfo  {journal} {Nature Communications}\ }\textbf
  {\bibinfo {volume} {10}} (\bibinfo {year} {2019}),\
  10.1038/s41467-019-10545-x}\BibitemShut {NoStop}%
\bibitem [{\citenamefont {Streib}\ \emph {et~al.}(2019)\citenamefont {Streib},
  \citenamefont {Vidal-Silva}, \citenamefont {Shen},\ and\ \citenamefont
  {Bauer}}]{streib_prb19}%
  \BibitemOpen
  \bibfield  {author} {\bibinfo {author} {\bibfnamefont {S.}~\bibnamefont
  {Streib}}, \bibinfo {author} {\bibfnamefont {N.}~\bibnamefont {Vidal-Silva}},
  \bibinfo {author} {\bibfnamefont {K.}~\bibnamefont {Shen}}, \ and\ \bibinfo
  {author} {\bibfnamefont {G.~E.}\ \bibnamefont {Bauer}},\ }\href {\doibase
  10.1103/PhysRevB.99.184442} {\bibfield  {journal} {\bibinfo  {journal} {Phys.
  Rev. B}\ }\textbf {\bibinfo {volume} {99}},\ \bibinfo {pages} {1} (\bibinfo
  {year} {2019})}\BibitemShut {NoStop}%
\bibitem [{\citenamefont {O’Connell}\ \emph {et~al.}(2010)\citenamefont
  {O’Connell}, \citenamefont {Hofheinz}, \citenamefont {Ansmann},
  \citenamefont {Bialczak}, \citenamefont {Lenander}, \citenamefont {Lucero},
  \citenamefont {Neeley}, \citenamefont {Sank}, \citenamefont {Wang},
  \citenamefont {Weides}, \citenamefont {Wenner}, \citenamefont {Martinis},\
  and\ \citenamefont {Cleland}}]{oconnell_nature10}%
  \BibitemOpen
  \bibfield  {author} {\bibinfo {author} {\bibfnamefont {A.~D.}\ \bibnamefont
  {O’Connell}}, \bibinfo {author} {\bibfnamefont {M.}~\bibnamefont
  {Hofheinz}}, \bibinfo {author} {\bibfnamefont {M.}~\bibnamefont {Ansmann}},
  \bibinfo {author} {\bibfnamefont {R.~C.}\ \bibnamefont {Bialczak}}, \bibinfo
  {author} {\bibfnamefont {M.}~\bibnamefont {Lenander}}, \bibinfo {author}
  {\bibfnamefont {E.}~\bibnamefont {Lucero}}, \bibinfo {author} {\bibfnamefont
  {M.}~\bibnamefont {Neeley}}, \bibinfo {author} {\bibfnamefont
  {D.}~\bibnamefont {Sank}}, \bibinfo {author} {\bibfnamefont {H.}~\bibnamefont
  {Wang}}, \bibinfo {author} {\bibfnamefont {M.}~\bibnamefont {Weides}},
  \bibinfo {author} {\bibfnamefont {J.}~\bibnamefont {Wenner}}, \bibinfo
  {author} {\bibfnamefont {J.~M.}\ \bibnamefont {Martinis}}, \ and\ \bibinfo
  {author} {\bibfnamefont {A.~N.}\ \bibnamefont {Cleland}},\ }\href {\doibase
  10.1038/nature08967} {\bibfield  {journal} {\bibinfo  {journal} {Nature}\
  }\textbf {\bibinfo {volume} {464}},\ \bibinfo {pages} {697} (\bibinfo {year}
  {2010})}\BibitemShut {NoStop}%
\bibitem [{\citenamefont {Chu}\ \emph {et~al.}(2017)\citenamefont {Chu},
  \citenamefont {Kharel}, \citenamefont {Renninger}, \citenamefont {Burkhart},
  \citenamefont {Frunzio}, \citenamefont {Rakich},\ and\ \citenamefont
  {Schoelkopf}}]{chu_science17}%
  \BibitemOpen
  \bibfield  {author} {\bibinfo {author} {\bibfnamefont {Y.}~\bibnamefont
  {Chu}}, \bibinfo {author} {\bibfnamefont {P.}~\bibnamefont {Kharel}},
  \bibinfo {author} {\bibfnamefont {W.~H.}\ \bibnamefont {Renninger}}, \bibinfo
  {author} {\bibfnamefont {L.~D.}\ \bibnamefont {Burkhart}}, \bibinfo {author}
  {\bibfnamefont {L.}~\bibnamefont {Frunzio}}, \bibinfo {author} {\bibfnamefont
  {P.~T.}\ \bibnamefont {Rakich}}, \ and\ \bibinfo {author} {\bibfnamefont
  {R.~J.}\ \bibnamefont {Schoelkopf}},\ }\href {\doibase
  10.1126/science.aao1511} {\bibfield  {journal} {\bibinfo  {journal}
  {Science}\ }\textbf {\bibinfo {volume} {358}},\ \bibinfo {pages} {199–202}
  (\bibinfo {year} {2017})}\BibitemShut {NoStop}%
\bibitem [{\citenamefont {Chu}\ \emph {et~al.}(2018)\citenamefont {Chu},
  \citenamefont {Kharel}, \citenamefont {Yoon}, \citenamefont {Frunzio},
  \citenamefont {Rakich},\ and\ \citenamefont {Schoelkopf}}]{chu_nature18}%
  \BibitemOpen
  \bibfield  {author} {\bibinfo {author} {\bibfnamefont {Y.}~\bibnamefont
  {Chu}}, \bibinfo {author} {\bibfnamefont {P.}~\bibnamefont {Kharel}},
  \bibinfo {author} {\bibfnamefont {T.}~\bibnamefont {Yoon}}, \bibinfo {author}
  {\bibfnamefont {L.}~\bibnamefont {Frunzio}}, \bibinfo {author} {\bibfnamefont
  {P.~T.}\ \bibnamefont {Rakich}}, \ and\ \bibinfo {author} {\bibfnamefont
  {R.~J.}\ \bibnamefont {Schoelkopf}},\ }\href {\doibase
  10.1038/s41586-018-0717-7} {\bibfield  {journal} {\bibinfo  {journal}
  {Nature}\ }\textbf {\bibinfo {volume} {563}},\ \bibinfo {pages} {666}
  (\bibinfo {year} {2018})}\BibitemShut {NoStop}%
\bibitem [{\citenamefont {Moores}\ \emph {et~al.}(2018)\citenamefont {Moores},
  \citenamefont {Sletten}, \citenamefont {Viennot},\ and\ \citenamefont
  {Lehnert}}]{moores_prl18}%
  \BibitemOpen
  \bibfield  {author} {\bibinfo {author} {\bibfnamefont {B.~A.}\ \bibnamefont
  {Moores}}, \bibinfo {author} {\bibfnamefont {L.~R.}\ \bibnamefont {Sletten}},
  \bibinfo {author} {\bibfnamefont {J.~J.}\ \bibnamefont {Viennot}}, \ and\
  \bibinfo {author} {\bibfnamefont {K.~W.}\ \bibnamefont {Lehnert}},\ }\href
  {\doibase 10.1103/PhysRevLett.120.227701} {\bibfield  {journal} {\bibinfo
  {journal} {Physical Review Letters}\ }\textbf {\bibinfo {volume} {120}},\
  \bibinfo {pages} {227701} (\bibinfo {year} {2018})}\BibitemShut {NoStop}%
\bibitem [{\citenamefont {Agnese}\ \emph {et~al.}(2014)\citenamefont {Agnese},
  \citenamefont {Anderson}, \citenamefont {Asai}, \citenamefont
  {Balakishiyeva}, \citenamefont {Basu~Thakur}, \citenamefont {Bauer},
  \citenamefont {Billard}, \citenamefont {Borgland}, \citenamefont {Bowles},
  \citenamefont {Brandt},\ and\ \citenamefont {et~al.}}]{cdmslite}%
  \BibitemOpen
  \bibfield  {author} {\bibinfo {author} {\bibfnamefont {R.}~\bibnamefont
  {Agnese}}, \bibinfo {author} {\bibfnamefont {A.~J.}\ \bibnamefont
  {Anderson}}, \bibinfo {author} {\bibfnamefont {M.}~\bibnamefont {Asai}},
  \bibinfo {author} {\bibfnamefont {D.}~\bibnamefont {Balakishiyeva}}, \bibinfo
  {author} {\bibfnamefont {R.}~\bibnamefont {Basu~Thakur}}, \bibinfo {author}
  {\bibfnamefont {D.~A.}\ \bibnamefont {Bauer}}, \bibinfo {author}
  {\bibfnamefont {J.}~\bibnamefont {Billard}}, \bibinfo {author} {\bibfnamefont
  {A.}~\bibnamefont {Borgland}}, \bibinfo {author} {\bibfnamefont {M.~A.}\
  \bibnamefont {Bowles}}, \bibinfo {author} {\bibfnamefont {D.}~\bibnamefont
  {Brandt}}, \ and\ \bibinfo {author} {\bibnamefont {et~al.}},\ }\href
  {\doibase 10.1103/physrevlett.112.041302} {\bibfield  {journal} {\bibinfo
  {journal} {Physical Review Letters}\ }\textbf {\bibinfo {volume} {112}}
  (\bibinfo {year} {2014}),\ 10.1103/physrevlett.112.041302}\BibitemShut
  {NoStop}%
\bibitem [{\citenamefont {Knapen}\ \emph {et~al.}(2017)\citenamefont {Knapen},
  \citenamefont {Lin},\ and\ \citenamefont {Zurek}}]{knapen_prd17_sfhelium}%
  \BibitemOpen
  \bibfield  {author} {\bibinfo {author} {\bibfnamefont {S.}~\bibnamefont
  {Knapen}}, \bibinfo {author} {\bibfnamefont {T.}~\bibnamefont {Lin}}, \ and\
  \bibinfo {author} {\bibfnamefont {K.~M.}\ \bibnamefont {Zurek}},\ }\href
  {\doibase 10.1103/PhysRevD.95.056019} {\bibfield  {journal} {\bibinfo
  {journal} {Physical Review D}\ }\textbf {\bibinfo {volume} {95}},\ \bibinfo
  {pages} {1} (\bibinfo {year} {2017})}\BibitemShut {NoStop}%
\bibitem [{\citenamefont {Griffin}\ \emph {et~al.}(2021)\citenamefont
  {Griffin}, \citenamefont {Hochberg}, \citenamefont {Inzani}, \citenamefont
  {Kurinsky}, \citenamefont {Lin},\ and\ \citenamefont {Yu}}]{griffin_21_sic}%
  \BibitemOpen
  \bibfield  {author} {\bibinfo {author} {\bibfnamefont {S.~M.}\ \bibnamefont
  {Griffin}}, \bibinfo {author} {\bibfnamefont {Y.}~\bibnamefont {Hochberg}},
  \bibinfo {author} {\bibfnamefont {K.}~\bibnamefont {Inzani}}, \bibinfo
  {author} {\bibfnamefont {N.}~\bibnamefont {Kurinsky}}, \bibinfo {author}
  {\bibfnamefont {T.}~\bibnamefont {Lin}}, \ and\ \bibinfo {author}
  {\bibfnamefont {T.~C.}\ \bibnamefont {Yu}},\ }\href {\doibase
  10.1103/physrevd.103.075002} {\bibfield  {journal} {\bibinfo  {journal}
  {Physical Review D}\ }\textbf {\bibinfo {volume} {103}} (\bibinfo {year}
  {2021}),\ 10.1103/physrevd.103.075002}\BibitemShut {NoStop}%
\bibitem [{\citenamefont {Griffin}\ \emph {et~al.}(2018)\citenamefont
  {Griffin}, \citenamefont {Knapen}, \citenamefont {Lin},\ and\ \citenamefont
  {Zurek}}]{griffin_prd18}%
  \BibitemOpen
  \bibfield  {author} {\bibinfo {author} {\bibfnamefont {S.}~\bibnamefont
  {Griffin}}, \bibinfo {author} {\bibfnamefont {S.}~\bibnamefont {Knapen}},
  \bibinfo {author} {\bibfnamefont {T.}~\bibnamefont {Lin}}, \ and\ \bibinfo
  {author} {\bibfnamefont {K.~M.}\ \bibnamefont {Zurek}},\ }\href {\doibase
  10.1103/PhysRevD.98.115034} {\bibfield  {journal} {\bibinfo  {journal}
  {Physical Review D}\ }\textbf {\bibinfo {volume} {98}},\ \bibinfo {pages}
  {115034} (\bibinfo {year} {2018})}\BibitemShut {NoStop}%
\bibitem [{\citenamefont {Kurinsky}\ \emph {et~al.}(2019)\citenamefont
  {Kurinsky}, \citenamefont {Yu}, \citenamefont {Hochberg},\ and\ \citenamefont
  {Cabrera}}]{kurinsky_19_diamond}%
  \BibitemOpen
  \bibfield  {author} {\bibinfo {author} {\bibfnamefont {N.}~\bibnamefont
  {Kurinsky}}, \bibinfo {author} {\bibfnamefont {T.~C.}\ \bibnamefont {Yu}},
  \bibinfo {author} {\bibfnamefont {Y.}~\bibnamefont {Hochberg}}, \ and\
  \bibinfo {author} {\bibfnamefont {B.}~\bibnamefont {Cabrera}},\ }\href
  {\doibase 10.1103/physrevd.99.123005} {\bibfield  {journal} {\bibinfo
  {journal} {Physical Review D}\ }\textbf {\bibinfo {volume} {99}} (\bibinfo
  {year} {2019}),\ 10.1103/physrevd.99.123005}\BibitemShut {NoStop}%
\bibitem [{\citenamefont {Griffin}\ \emph {et~al.}(2020)\citenamefont
  {Griffin}, \citenamefont {Inzani}, \citenamefont {Trickle}, \citenamefont
  {Zhang},\ and\ \citenamefont {Zurek}}]{griffin_prd20}%
  \BibitemOpen
  \bibfield  {author} {\bibinfo {author} {\bibfnamefont {S.~M.}\ \bibnamefont
  {Griffin}}, \bibinfo {author} {\bibfnamefont {K.}~\bibnamefont {Inzani}},
  \bibinfo {author} {\bibfnamefont {T.}~\bibnamefont {Trickle}}, \bibinfo
  {author} {\bibfnamefont {Z.}~\bibnamefont {Zhang}}, \ and\ \bibinfo {author}
  {\bibfnamefont {K.~M.}\ \bibnamefont {Zurek}},\ }\href {\doibase
  10.1103/PhysRevD.101.055004} {\bibfield  {journal} {\bibinfo  {journal}
  {Physical Review D}\ }\textbf {\bibinfo {volume} {101}},\ \bibinfo {pages}
  {1} (\bibinfo {year} {2020})}\BibitemShut {NoStop}%
\bibitem [{\citenamefont {Trickle}\ \emph {et~al.}(2020)\citenamefont
  {Trickle}, \citenamefont {Zhang},\ and\ \citenamefont
  {Zurek}}]{trickle_prl20}%
  \BibitemOpen
  \bibfield  {author} {\bibinfo {author} {\bibfnamefont {T.}~\bibnamefont
  {Trickle}}, \bibinfo {author} {\bibfnamefont {Z.}~\bibnamefont {Zhang}}, \
  and\ \bibinfo {author} {\bibfnamefont {K.~M.}\ \bibnamefont {Zurek}},\ }\href
  {\doibase 10.1103/PhysRevLett.124.201801} {\bibfield  {journal} {\bibinfo
  {journal} {Physical Review Letters}\ }\textbf {\bibinfo {volume} {124}},\
  \bibinfo {pages} {1} (\bibinfo {year} {2020})}\BibitemShut {NoStop}%
\bibitem [{\citenamefont {Hochberg}\ \emph {et~al.}(2018)\citenamefont
  {Hochberg}, \citenamefont {Kahn}, \citenamefont {Lisanti}, \citenamefont
  {Zurek}, \citenamefont {Grushin}, \citenamefont {Ilan}, \citenamefont
  {Griffin}, \citenamefont {Liu}, \citenamefont {Weber},\ and\ \citenamefont
  {Neaton}}]{hochberg_prd18_dirac}%
  \BibitemOpen
  \bibfield  {author} {\bibinfo {author} {\bibfnamefont {Y.}~\bibnamefont
  {Hochberg}}, \bibinfo {author} {\bibfnamefont {Y.}~\bibnamefont {Kahn}},
  \bibinfo {author} {\bibfnamefont {M.}~\bibnamefont {Lisanti}}, \bibinfo
  {author} {\bibfnamefont {K.~M.}\ \bibnamefont {Zurek}}, \bibinfo {author}
  {\bibfnamefont {A.~G.}\ \bibnamefont {Grushin}}, \bibinfo {author}
  {\bibfnamefont {R.}~\bibnamefont {Ilan}}, \bibinfo {author} {\bibfnamefont
  {S.~M.}\ \bibnamefont {Griffin}}, \bibinfo {author} {\bibfnamefont {Z.-F.}\
  \bibnamefont {Liu}}, \bibinfo {author} {\bibfnamefont {S.~F.}\ \bibnamefont
  {Weber}}, \ and\ \bibinfo {author} {\bibfnamefont {J.~B.}\ \bibnamefont
  {Neaton}},\ }\href {\doibase 10.1103/PhysRevD.97.015004} {\bibfield
  {journal} {\bibinfo  {journal} {Physical Review D}\ }\textbf {\bibinfo
  {volume} {97}},\ \bibinfo {pages} {015004} (\bibinfo {year}
  {2018})}\BibitemShut {NoStop}%
\bibitem [{\citenamefont {Røising}\ \emph {et~al.}(2021)\citenamefont
  {Røising}, \citenamefont {Fraser}, \citenamefont {Griffin}, \citenamefont
  {Bandyopadhyay}, \citenamefont {Mahabir}, \citenamefont {Cheong},\ and\
  \citenamefont {Balatsky}}]{Roising_et_al:2021}%
  \BibitemOpen
  \bibfield  {author} {\bibinfo {author} {\bibfnamefont {H.~S.}\ \bibnamefont
  {Røising}}, \bibinfo {author} {\bibfnamefont {B.}~\bibnamefont {Fraser}},
  \bibinfo {author} {\bibfnamefont {S.~M.}\ \bibnamefont {Griffin}}, \bibinfo
  {author} {\bibfnamefont {S.}~\bibnamefont {Bandyopadhyay}}, \bibinfo {author}
  {\bibfnamefont {A.}~\bibnamefont {Mahabir}}, \bibinfo {author} {\bibfnamefont
  {S.-W.}\ \bibnamefont {Cheong}}, \ and\ \bibinfo {author} {\bibfnamefont
  {A.~V.}\ \bibnamefont {Balatsky}},\ }\href {\doibase
  10.1103/physrevresearch.3.033236} {\bibfield  {journal} {\bibinfo  {journal}
  {Physical Review Research}\ }\textbf {\bibinfo {volume} {3}} (\bibinfo {year}
  {2021}),\ 10.1103/physrevresearch.3.033236}\BibitemShut {NoStop}%
\bibitem [{\citenamefont {Inzani}\ \emph {et~al.}(2021)\citenamefont {Inzani},
  \citenamefont {Faghaninia},\ and\ \citenamefont
  {Griffin}}]{Inzani_et_al:2021}%
  \BibitemOpen
  \bibfield  {author} {\bibinfo {author} {\bibfnamefont {K.}~\bibnamefont
  {Inzani}}, \bibinfo {author} {\bibfnamefont {A.}~\bibnamefont {Faghaninia}},
  \ and\ \bibinfo {author} {\bibfnamefont {S.~M.}\ \bibnamefont {Griffin}},\
  }\href@noop {} {\bibfield  {journal} {\bibinfo  {journal} {Physical Review
  Research}\ }\textbf {\bibinfo {volume} {3}},\ \bibinfo {pages} {013069}
  (\bibinfo {year} {2021})}\BibitemShut {NoStop}%
\bibitem [{\citenamefont {Zmuidzinas}(2012)}]{zmuidzinas_annrevcmp12}%
  \BibitemOpen
  \bibfield  {author} {\bibinfo {author} {\bibfnamefont {J.}~\bibnamefont
  {Zmuidzinas}},\ }\href {\doibase 10.1146/annurev-conmatphys-020911-125022}
  {\bibfield  {journal} {\bibinfo  {journal} {Annual Review of Condensed Matter
  Physics}\ }\textbf {\bibinfo {volume} {3}},\ \bibinfo {pages} {169} (\bibinfo
  {year} {2012})}\BibitemShut {NoStop}%
\bibitem [{\citenamefont {Fleischmann}\ \emph {et~al.}(2005)\citenamefont
  {Fleischmann}, \citenamefont {Enss},\ and\ \citenamefont
  {Seidel}}]{Fleischmann2005MetallicCalorimeters}%
  \BibitemOpen
  \bibfield  {author} {\bibinfo {author} {\bibfnamefont {A.}~\bibnamefont
  {Fleischmann}}, \bibinfo {author} {\bibfnamefont {C.}~\bibnamefont {Enss}}, \
  and\ \bibinfo {author} {\bibfnamefont {G.}~\bibnamefont {Seidel}},\ }\enquote
  {\bibinfo {title} {Metallic magnetic calorimeters},}\ in\ \href {\doibase
  10.1007/10933596_4} {\emph {\bibinfo {booktitle} {Cryogenic Particle
  Detection}}},\ \bibinfo {editor} {edited by\ \bibinfo {editor} {\bibfnamefont
  {C.}~\bibnamefont {Enss}}}\ (\bibinfo  {publisher} {Springer Berlin
  Heidelberg},\ \bibinfo {address} {Berlin, Heidelberg},\ \bibinfo {year}
  {2005})\ pp.\ \bibinfo {pages} {151--216}\BibitemShut {NoStop}%
\bibitem [{\citenamefont {Irwin}\ \emph {et~al.}(1995)\citenamefont {Irwin},
  \citenamefont {Nam}, \citenamefont {Cabrera}, \citenamefont {Chugg},\ and\
  \citenamefont {Young}}]{tes_review}%
  \BibitemOpen
  \bibfield  {author} {\bibinfo {author} {\bibfnamefont {K.~D.}\ \bibnamefont
  {Irwin}}, \bibinfo {author} {\bibfnamefont {S.~W.}\ \bibnamefont {Nam}},
  \bibinfo {author} {\bibfnamefont {B.}~\bibnamefont {Cabrera}}, \bibinfo
  {author} {\bibfnamefont {B.}~\bibnamefont {Chugg}}, \ and\ \bibinfo {author}
  {\bibfnamefont {B.~A.}\ \bibnamefont {Young}},\ }\href {\doibase
  10.1063/1.1146105} {\bibfield  {journal} {\bibinfo  {journal} {Review of
  Scientific Instruments}\ }\textbf {\bibinfo {volume} {66}},\ \bibinfo {pages}
  {5322} (\bibinfo {year} {1995})}\BibitemShut {NoStop}%
\bibitem [{\citenamefont {Romani}\ \emph {et~al.}(2018)\citenamefont {Romani},
  \citenamefont {Brink}, \citenamefont {Cabrera}, \citenamefont {Cherry},
  \citenamefont {Howarth}, \citenamefont {Kurinsky}, \citenamefont {Moffatt},
  \citenamefont {Partridge}, \citenamefont {Ponce}, \citenamefont {Pyle},
  \citenamefont {Tomada}, \citenamefont {Yellin}, \citenamefont {Yen},\ and\
  \citenamefont {Young}}]{pyle_apl18_tes}%
  \BibitemOpen
  \bibfield  {author} {\bibinfo {author} {\bibfnamefont {R.~K.}\ \bibnamefont
  {Romani}}, \bibinfo {author} {\bibfnamefont {P.~L.}\ \bibnamefont {Brink}},
  \bibinfo {author} {\bibfnamefont {B.}~\bibnamefont {Cabrera}}, \bibinfo
  {author} {\bibfnamefont {M.}~\bibnamefont {Cherry}}, \bibinfo {author}
  {\bibfnamefont {T.}~\bibnamefont {Howarth}}, \bibinfo {author} {\bibfnamefont
  {N.}~\bibnamefont {Kurinsky}}, \bibinfo {author} {\bibfnamefont {R.~A.}\
  \bibnamefont {Moffatt}}, \bibinfo {author} {\bibfnamefont {R.}~\bibnamefont
  {Partridge}}, \bibinfo {author} {\bibfnamefont {F.}~\bibnamefont {Ponce}},
  \bibinfo {author} {\bibfnamefont {M.}~\bibnamefont {Pyle}}, \bibinfo {author}
  {\bibfnamefont {A.}~\bibnamefont {Tomada}}, \bibinfo {author} {\bibfnamefont
  {S.}~\bibnamefont {Yellin}}, \bibinfo {author} {\bibfnamefont {J.~J.}\
  \bibnamefont {Yen}}, \ and\ \bibinfo {author} {\bibfnamefont {B.~A.}\
  \bibnamefont {Young}},\ }\href {\doibase 10.1063/1.5010699} {\bibfield
  {journal} {\bibinfo  {journal} {Applied Physics Letters}\ }\textbf {\bibinfo
  {volume} {112}} (\bibinfo {year} {2018}),\ 10.1063/1.5010699}\BibitemShut
  {NoStop}%
\bibitem [{\citenamefont {Angloher}\ \emph {et~al.}(2016)\citenamefont
  {Angloher}, \citenamefont {Bento}, \citenamefont {Bucci}, \citenamefont
  {Canonica}, \citenamefont {Defay}, \citenamefont {Erb}, \citenamefont {von
  Feilitzsch}, \citenamefont {Iachellini}, \citenamefont {Gorla}, \citenamefont
  {G{\"{u}}tlein}, \citenamefont {Hauff}, \citenamefont {Jochum}, \citenamefont
  {Kiefer}, \citenamefont {Kluck}, \citenamefont {Kraus}, \citenamefont
  {Lanfranchi}, \citenamefont {Loebell}, \citenamefont {M{\"{u}}nster},
  \citenamefont {Pagliarone}, \citenamefont {Petricca}, \citenamefont {Potzel},
  \citenamefont {Pr{\"{o}}bst}, \citenamefont {Reindl}, \citenamefont
  {Sch{\"{a}}ffner}, \citenamefont {Schieck}, \citenamefont {Sch{\"{o}}nert},
  \citenamefont {Seidel}, \citenamefont {Stodolsky}, \citenamefont
  {Strandhagen}, \citenamefont {Strauss}, \citenamefont {Tanzke}, \citenamefont
  {Trinh~Thi}, \citenamefont {T{\"{u}}rko{\u{g}}lu}, \citenamefont {Uffinger},
  \citenamefont {Ulrich}, \citenamefont {Usherov}, \citenamefont {Wawoczny},
  \citenamefont {Willers}, \citenamefont {W{\"{u}}strich},\ and\ \citenamefont
  {Z{\"{o}}ller}}]{cresst_ii}%
  \BibitemOpen
  \bibfield  {author} {\bibinfo {author} {\bibfnamefont {G.}~\bibnamefont
  {Angloher}}, \bibinfo {author} {\bibfnamefont {A.}~\bibnamefont {Bento}},
  \bibinfo {author} {\bibfnamefont {C.}~\bibnamefont {Bucci}}, \bibinfo
  {author} {\bibfnamefont {L.}~\bibnamefont {Canonica}}, \bibinfo {author}
  {\bibfnamefont {X.}~\bibnamefont {Defay}}, \bibinfo {author} {\bibfnamefont
  {A.}~\bibnamefont {Erb}}, \bibinfo {author} {\bibfnamefont {F.}~\bibnamefont
  {von Feilitzsch}}, \bibinfo {author} {\bibfnamefont {N.~F.}\ \bibnamefont
  {Iachellini}}, \bibinfo {author} {\bibfnamefont {P.}~\bibnamefont {Gorla}},
  \bibinfo {author} {\bibfnamefont {A.}~\bibnamefont {G{\"{u}}tlein}}, \bibinfo
  {author} {\bibfnamefont {D.}~\bibnamefont {Hauff}}, \bibinfo {author}
  {\bibfnamefont {J.}~\bibnamefont {Jochum}}, \bibinfo {author} {\bibfnamefont
  {M.}~\bibnamefont {Kiefer}}, \bibinfo {author} {\bibfnamefont
  {H.}~\bibnamefont {Kluck}}, \bibinfo {author} {\bibfnamefont
  {H.}~\bibnamefont {Kraus}}, \bibinfo {author} {\bibfnamefont {J.~C.}\
  \bibnamefont {Lanfranchi}}, \bibinfo {author} {\bibfnamefont
  {J.}~\bibnamefont {Loebell}}, \bibinfo {author} {\bibfnamefont
  {A.}~\bibnamefont {M{\"{u}}nster}}, \bibinfo {author} {\bibfnamefont
  {C.}~\bibnamefont {Pagliarone}}, \bibinfo {author} {\bibfnamefont
  {F.}~\bibnamefont {Petricca}}, \bibinfo {author} {\bibfnamefont
  {W.}~\bibnamefont {Potzel}}, \bibinfo {author} {\bibfnamefont
  {F.}~\bibnamefont {Pr{\"{o}}bst}}, \bibinfo {author} {\bibfnamefont
  {F.}~\bibnamefont {Reindl}}, \bibinfo {author} {\bibfnamefont
  {K.}~\bibnamefont {Sch{\"{a}}ffner}}, \bibinfo {author} {\bibfnamefont
  {J.}~\bibnamefont {Schieck}}, \bibinfo {author} {\bibfnamefont
  {S.}~\bibnamefont {Sch{\"{o}}nert}}, \bibinfo {author} {\bibfnamefont
  {W.}~\bibnamefont {Seidel}}, \bibinfo {author} {\bibfnamefont
  {L.}~\bibnamefont {Stodolsky}}, \bibinfo {author} {\bibfnamefont
  {C.}~\bibnamefont {Strandhagen}}, \bibinfo {author} {\bibfnamefont
  {R.}~\bibnamefont {Strauss}}, \bibinfo {author} {\bibfnamefont
  {A.}~\bibnamefont {Tanzke}}, \bibinfo {author} {\bibfnamefont {H.~H.}\
  \bibnamefont {Trinh~Thi}}, \bibinfo {author} {\bibfnamefont {C.}~\bibnamefont
  {T{\"{u}}rko{\u{g}}lu}}, \bibinfo {author} {\bibfnamefont {M.}~\bibnamefont
  {Uffinger}}, \bibinfo {author} {\bibfnamefont {A.}~\bibnamefont {Ulrich}},
  \bibinfo {author} {\bibfnamefont {I.}~\bibnamefont {Usherov}}, \bibinfo
  {author} {\bibfnamefont {S.}~\bibnamefont {Wawoczny}}, \bibinfo {author}
  {\bibfnamefont {M.}~\bibnamefont {Willers}}, \bibinfo {author} {\bibfnamefont
  {M.}~\bibnamefont {W{\"{u}}strich}}, \ and\ \bibinfo {author} {\bibfnamefont
  {A.}~\bibnamefont {Z{\"{o}}ller}},\ }\href {\doibase
  10.1140/epjc/s10052-016-3877-3} {\bibfield  {journal} {\bibinfo  {journal}
  {The European Physical Journal C}\ }\textbf {\bibinfo {volume} {76}},\
  \bibinfo {pages} {25} (\bibinfo {year} {2016})}\BibitemShut {NoStop}%
\bibitem [{\citenamefont {Petricca}\ \emph {et~al.}(2020)\citenamefont
  {Petricca}, \citenamefont {Angloher}, \citenamefont {Bauer}, \citenamefont
  {Bento}, \citenamefont {Bucci}, \citenamefont {Canonica}, \citenamefont
  {Defay}, \citenamefont {Erb}, \citenamefont {Feilitzsch}, \citenamefont
  {Iachellini}, \citenamefont {Gorla}, \citenamefont {G{\"{u}}tlein},
  \citenamefont {Hauff}, \citenamefont {Jochum}, \citenamefont {Kiefer},
  \citenamefont {Kluck}, \citenamefont {Kraus}, \citenamefont {Lanfranchi},
  \citenamefont {Langenk{\"{a}}mper}, \citenamefont {Loebell}, \citenamefont
  {Mancuso}, \citenamefont {Mondragon}, \citenamefont {M{\"{u}}nster},
  \citenamefont {Pagliarone}, \citenamefont {Potzel}, \citenamefont
  {Pr{\"{o}}bst}, \citenamefont {Puig}, \citenamefont {Reindl}, \citenamefont
  {Rothe}, \citenamefont {Sch{\"{a}}ffner}, \citenamefont {Schieck},
  \citenamefont {Sch{\"{o}}nert}, \citenamefont {Seidelf}, \citenamefont
  {Stahlberg}, \citenamefont {Stodolsky}, \citenamefont {Strandhagen},
  \citenamefont {Strauss}, \citenamefont {Tanzke}, \citenamefont {Thi},
  \citenamefont {T{\"{u}}rko{\u{g}}lu}, \citenamefont {Ulrich}, \citenamefont
  {Usherov}, \citenamefont {Wawoczny}, \citenamefont {Willers},\ and\
  \citenamefont {W{\"{u}}strich}}]{cresst_iii}%
  \BibitemOpen
  \bibfield  {author} {\bibinfo {author} {\bibfnamefont {F.}~\bibnamefont
  {Petricca}}, \bibinfo {author} {\bibfnamefont {G.}~\bibnamefont {Angloher}},
  \bibinfo {author} {\bibfnamefont {P.}~\bibnamefont {Bauer}}, \bibinfo
  {author} {\bibfnamefont {A.}~\bibnamefont {Bento}}, \bibinfo {author}
  {\bibfnamefont {C.}~\bibnamefont {Bucci}}, \bibinfo {author} {\bibfnamefont
  {L.}~\bibnamefont {Canonica}}, \bibinfo {author} {\bibfnamefont
  {X.}~\bibnamefont {Defay}}, \bibinfo {author} {\bibfnamefont
  {A.}~\bibnamefont {Erb}}, \bibinfo {author} {\bibfnamefont {F.~v.}\
  \bibnamefont {Feilitzsch}}, \bibinfo {author} {\bibfnamefont {N.~F.}\
  \bibnamefont {Iachellini}}, \bibinfo {author} {\bibfnamefont
  {P.}~\bibnamefont {Gorla}}, \bibinfo {author} {\bibfnamefont
  {A.}~\bibnamefont {G{\"{u}}tlein}}, \bibinfo {author} {\bibfnamefont
  {D.}~\bibnamefont {Hauff}}, \bibinfo {author} {\bibfnamefont
  {J.}~\bibnamefont {Jochum}}, \bibinfo {author} {\bibfnamefont
  {M.}~\bibnamefont {Kiefer}}, \bibinfo {author} {\bibfnamefont
  {H.}~\bibnamefont {Kluck}}, \bibinfo {author} {\bibfnamefont
  {H.}~\bibnamefont {Kraus}}, \bibinfo {author} {\bibfnamefont {J.~C.}\
  \bibnamefont {Lanfranchi}}, \bibinfo {author} {\bibfnamefont
  {A.}~\bibnamefont {Langenk{\"{a}}mper}}, \bibinfo {author} {\bibfnamefont
  {J.}~\bibnamefont {Loebell}}, \bibinfo {author} {\bibfnamefont
  {M.}~\bibnamefont {Mancuso}}, \bibinfo {author} {\bibfnamefont
  {E.}~\bibnamefont {Mondragon}}, \bibinfo {author} {\bibfnamefont
  {A.}~\bibnamefont {M{\"{u}}nster}}, \bibinfo {author} {\bibfnamefont
  {C.}~\bibnamefont {Pagliarone}}, \bibinfo {author} {\bibfnamefont
  {W.}~\bibnamefont {Potzel}}, \bibinfo {author} {\bibfnamefont
  {F.}~\bibnamefont {Pr{\"{o}}bst}}, \bibinfo {author} {\bibfnamefont
  {R.}~\bibnamefont {Puig}}, \bibinfo {author} {\bibfnamefont {F.}~\bibnamefont
  {Reindl}}, \bibinfo {author} {\bibfnamefont {J.}~\bibnamefont {Rothe}},
  \bibinfo {author} {\bibfnamefont {K.}~\bibnamefont {Sch{\"{a}}ffner}},
  \bibinfo {author} {\bibfnamefont {J.}~\bibnamefont {Schieck}}, \bibinfo
  {author} {\bibfnamefont {S.}~\bibnamefont {Sch{\"{o}}nert}}, \bibinfo
  {author} {\bibfnamefont {W.}~\bibnamefont {Seidelf}}, \bibinfo {author}
  {\bibfnamefont {M.}~\bibnamefont {Stahlberg}}, \bibinfo {author}
  {\bibfnamefont {L.}~\bibnamefont {Stodolsky}}, \bibinfo {author}
  {\bibfnamefont {C.}~\bibnamefont {Strandhagen}}, \bibinfo {author}
  {\bibfnamefont {R.}~\bibnamefont {Strauss}}, \bibinfo {author} {\bibfnamefont
  {A.}~\bibnamefont {Tanzke}}, \bibinfo {author} {\bibfnamefont {H.~H.~T.}\
  \bibnamefont {Thi}}, \bibinfo {author} {\bibfnamefont {C.}~\bibnamefont
  {T{\"{u}}rko{\u{g}}lu}}, \bibinfo {author} {\bibfnamefont {A.}~\bibnamefont
  {Ulrich}}, \bibinfo {author} {\bibfnamefont {I.}~\bibnamefont {Usherov}},
  \bibinfo {author} {\bibfnamefont {S.}~\bibnamefont {Wawoczny}}, \bibinfo
  {author} {\bibfnamefont {M.}~\bibnamefont {Willers}}, \ and\ \bibinfo
  {author} {\bibfnamefont {M.}~\bibnamefont {W{\"{u}}strich}},\ }\href
  {\doibase 10.1088/1742-6596/1342/1/012076} {\bibfield  {journal} {\bibinfo
  {journal} {Journal of Physics: Conference Series}\ }\textbf {\bibinfo
  {volume} {1342}},\ \bibinfo {pages} {012076} (\bibinfo {year}
  {2020})}\BibitemShut {NoStop}%
\bibitem [{\citenamefont {Day}\ \emph {et~al.}(2003)\citenamefont {Day},
  \citenamefont {LeDuc}, \citenamefont {Mazin}, \citenamefont {Vayonakis},\
  and\ \citenamefont {Zmuidzinas}}]{zmuidzinas_nature}%
  \BibitemOpen
  \bibfield  {author} {\bibinfo {author} {\bibfnamefont {P.~K.}\ \bibnamefont
  {Day}}, \bibinfo {author} {\bibfnamefont {H.~G.}\ \bibnamefont {LeDuc}},
  \bibinfo {author} {\bibfnamefont {B.~A.}\ \bibnamefont {Mazin}}, \bibinfo
  {author} {\bibfnamefont {A.}~\bibnamefont {Vayonakis}}, \ and\ \bibinfo
  {author} {\bibfnamefont {J.}~\bibnamefont {Zmuidzinas}},\ }\href {\doibase
  10.1038/nature02037} {\bibfield  {journal} {\bibinfo  {journal} {Nature}\
  }\textbf {\bibinfo {volume} {425}},\ \bibinfo {pages} {817} (\bibinfo {year}
  {2003})}\BibitemShut {NoStop}%
\bibitem [{\citenamefont {Fink}\ \emph {et~al.}(2020)\citenamefont {Fink},
  \citenamefont {Watkins}, \citenamefont {Aramaki}, \citenamefont {Brink},
  \citenamefont {Ganjam}, \citenamefont {Hines}, \citenamefont {Huber},
  \citenamefont {Kurinsky}, \citenamefont {Mahapatra}, \citenamefont
  {Mirabolfathi}, \citenamefont {Page}, \citenamefont {Partridge},
  \citenamefont {Platt}, \citenamefont {Pyle}, \citenamefont {Sadoulet},
  \citenamefont {Serfass},\ and\ \citenamefont {Zuber}}]{fink_aip20}%
  \BibitemOpen
  \bibfield  {author} {\bibinfo {author} {\bibfnamefont {C.~W.}\ \bibnamefont
  {Fink}}, \bibinfo {author} {\bibfnamefont {S.~L.}\ \bibnamefont {Watkins}},
  \bibinfo {author} {\bibfnamefont {T.}~\bibnamefont {Aramaki}}, \bibinfo
  {author} {\bibfnamefont {P.~L.}\ \bibnamefont {Brink}}, \bibinfo {author}
  {\bibfnamefont {S.}~\bibnamefont {Ganjam}}, \bibinfo {author} {\bibfnamefont
  {B.~A.}\ \bibnamefont {Hines}}, \bibinfo {author} {\bibfnamefont {M.~E.}\
  \bibnamefont {Huber}}, \bibinfo {author} {\bibfnamefont {N.~A.}\ \bibnamefont
  {Kurinsky}}, \bibinfo {author} {\bibfnamefont {R.}~\bibnamefont {Mahapatra}},
  \bibinfo {author} {\bibfnamefont {N.}~\bibnamefont {Mirabolfathi}}, \bibinfo
  {author} {\bibfnamefont {W.~A.}\ \bibnamefont {Page}}, \bibinfo {author}
  {\bibfnamefont {R.}~\bibnamefont {Partridge}}, \bibinfo {author}
  {\bibfnamefont {M.}~\bibnamefont {Platt}}, \bibinfo {author} {\bibfnamefont
  {M.}~\bibnamefont {Pyle}}, \bibinfo {author} {\bibfnamefont {B.}~\bibnamefont
  {Sadoulet}}, \bibinfo {author} {\bibfnamefont {B.}~\bibnamefont {Serfass}}, \
  and\ \bibinfo {author} {\bibfnamefont {S.}~\bibnamefont {Zuber}},\ }\href
  {\doibase 10.1063/5.0011130} {\bibfield  {journal} {\bibinfo  {journal} {AIP
  Advances}\ }\textbf {\bibinfo {volume} {10}},\ \bibinfo {pages} {085221}
  (\bibinfo {year} {2020})}\BibitemShut {NoStop}%
\bibitem [{\citenamefont {Arkani-Hamed}\ and\ \citenamefont
  {Weiner}(2008)}]{arkani_jhep08}%
  \BibitemOpen
  \bibfield  {author} {\bibinfo {author} {\bibfnamefont {N.}~\bibnamefont
  {Arkani-Hamed}}\ and\ \bibinfo {author} {\bibfnamefont {N.}~\bibnamefont
  {Weiner}},\ }\href {\doibase 10.1088/1126-6708/2008/12/104} {\bibfield
  {journal} {\bibinfo  {journal} {Journal of High Energy Physics}\ }\textbf
  {\bibinfo {volume} {2008}},\ \bibinfo {pages} {104} (\bibinfo {year}
  {2008})}\BibitemShut {NoStop}%
\bibitem [{\citenamefont {Cheung}\ \emph {et~al.}(2009)\citenamefont {Cheung},
  \citenamefont {Ruderman}, \citenamefont {Wang},\ and\ \citenamefont
  {Yavin}}]{cheung_prd09}%
  \BibitemOpen
  \bibfield  {author} {\bibinfo {author} {\bibfnamefont {C.}~\bibnamefont
  {Cheung}}, \bibinfo {author} {\bibfnamefont {J.~T.}\ \bibnamefont
  {Ruderman}}, \bibinfo {author} {\bibfnamefont {L.-T.}\ \bibnamefont {Wang}},
  \ and\ \bibinfo {author} {\bibfnamefont {I.}~\bibnamefont {Yavin}},\ }\href
  {\doibase 10.1103/PhysRevD.80.035008} {\bibfield  {journal} {\bibinfo
  {journal} {Physical Review D}\ }\textbf {\bibinfo {volume} {80}},\ \bibinfo
  {pages} {035008} (\bibinfo {year} {2009})}\BibitemShut {NoStop}%
\bibitem [{\citenamefont {Morrissey}\ \emph {et~al.}(2009)\citenamefont
  {Morrissey}, \citenamefont {Poland},\ and\ \citenamefont
  {Zurek}}]{morrissey_jhep09}%
  \BibitemOpen
  \bibfield  {author} {\bibinfo {author} {\bibfnamefont {D.~E.}\ \bibnamefont
  {Morrissey}}, \bibinfo {author} {\bibfnamefont {D.}~\bibnamefont {Poland}}, \
  and\ \bibinfo {author} {\bibfnamefont {K.~M.}\ \bibnamefont {Zurek}},\ }\href
  {\doibase 10.1088/1126-6708/2009/07/050} {\bibfield  {journal} {\bibinfo
  {journal} {Journal of High Energy Physics}\ }\textbf {\bibinfo {volume}
  {2009}},\ \bibinfo {pages} {050} (\bibinfo {year} {2009})}\BibitemShut
  {NoStop}%
\bibitem [{\citenamefont {Fabbrichesi}\ \emph {et~al.}(2021)\citenamefont
  {Fabbrichesi}, \citenamefont {Gabrielli},\ and\ \citenamefont
  {Lanfranchi}}]{darkphoton}%
  \BibitemOpen
  \bibfield  {author} {\bibinfo {author} {\bibfnamefont {M.}~\bibnamefont
  {Fabbrichesi}}, \bibinfo {author} {\bibfnamefont {E.}~\bibnamefont
  {Gabrielli}}, \ and\ \bibinfo {author} {\bibfnamefont {G.}~\bibnamefont
  {Lanfranchi}},\ }\href {\doibase 10.1007/978-3-030-62519-1} {\emph {\bibinfo
  {title} {{The Physics of the Dark Photon}}}},\ SpringerBriefs in Physics\
  (\bibinfo  {publisher} {Springer International Publishing},\ \bibinfo
  {address} {Cham},\ \bibinfo {year} {2021})\BibitemShut {NoStop}%
\bibitem [{\citenamefont {Hohenberg}\ and\ \citenamefont
  {Kohn}(1964)}]{HohenbergKohn1964}%
  \BibitemOpen
  \bibfield  {author} {\bibinfo {author} {\bibfnamefont {P.}~\bibnamefont
  {Hohenberg}}\ and\ \bibinfo {author} {\bibfnamefont {W.}~\bibnamefont
  {Kohn}},\ }\href {\doibase 10.1103/PhysRev.136.B864} {\bibfield  {journal}
  {\bibinfo  {journal} {Phys. Rev.}\ }\textbf {\bibinfo {volume} {136}},\
  \bibinfo {pages} {B864} (\bibinfo {year} {1964})}\BibitemShut {NoStop}%
\bibitem [{\citenamefont {Kohn}\ and\ \citenamefont
  {Sham}(1965)}]{KohnSham1965}%
  \BibitemOpen
  \bibfield  {author} {\bibinfo {author} {\bibfnamefont {W.}~\bibnamefont
  {Kohn}}\ and\ \bibinfo {author} {\bibfnamefont {L.~J.}\ \bibnamefont
  {Sham}},\ }\href {\doibase 10.1103/PhysRev.140.A1133} {\bibfield  {journal}
  {\bibinfo  {journal} {Phys. Rev.}\ }\textbf {\bibinfo {volume} {140}},\
  \bibinfo {pages} {A1133} (\bibinfo {year} {1965})}\BibitemShut {NoStop}%
\bibitem [{\citenamefont {Togo}\ \emph {et~al.}(2015)\citenamefont {Togo},
  \citenamefont {Chaput},\ and\ \citenamefont {Tanaka}}]{phono3py}%
  \BibitemOpen
  \bibfield  {author} {\bibinfo {author} {\bibfnamefont {A.}~\bibnamefont
  {Togo}}, \bibinfo {author} {\bibfnamefont {L.}~\bibnamefont {Chaput}}, \ and\
  \bibinfo {author} {\bibfnamefont {I.}~\bibnamefont {Tanaka}},\ }\href
  {\doibase 10.1103/PhysRevB.91.094306} {\bibfield  {journal} {\bibinfo
  {journal} {Phys. Rev. B}\ }\textbf {\bibinfo {volume} {91}} (\bibinfo {year}
  {2015}),\ 10.1103/PhysRevB.91.094306}\BibitemShut {NoStop}%
\bibitem [{\citenamefont {Tamura}(1983)}]{Tamura1983IsotopeGe}%
  \BibitemOpen
  \bibfield  {author} {\bibinfo {author} {\bibfnamefont {S.-i.}\ \bibnamefont
  {Tamura}},\ }\href {\doibase 10.1103/PhysRevB.27.858} {\bibfield  {journal}
  {\bibinfo  {journal} {Phys. Rev. B}\ }\textbf {\bibinfo {volume} {27}},\
  \bibinfo {pages} {858} (\bibinfo {year} {1983})}\BibitemShut {NoStop}%
\bibitem [{\citenamefont {Lindsay}\ \emph {et~al.}(2014)\citenamefont
  {Lindsay}, \citenamefont {Li}, \citenamefont {Carrete}, \citenamefont
  {Mingo}, \citenamefont {Broido},\ and\ \citenamefont
  {Reinecke}}]{lindsay_prb14}%
  \BibitemOpen
  \bibfield  {author} {\bibinfo {author} {\bibfnamefont {L.}~\bibnamefont
  {Lindsay}}, \bibinfo {author} {\bibfnamefont {W.}~\bibnamefont {Li}},
  \bibinfo {author} {\bibfnamefont {J.}~\bibnamefont {Carrete}}, \bibinfo
  {author} {\bibfnamefont {N.}~\bibnamefont {Mingo}}, \bibinfo {author}
  {\bibfnamefont {D.~A.}\ \bibnamefont {Broido}}, \ and\ \bibinfo {author}
  {\bibfnamefont {T.~L.}\ \bibnamefont {Reinecke}},\ }\href {\doibase
  10.1103/PhysRevB.89.155426} {\bibfield  {journal} {\bibinfo  {journal} {Phys.
  Rev. B}\ }\textbf {\bibinfo {volume} {89}},\ \bibinfo {pages} {155426}
  (\bibinfo {year} {2014})}\BibitemShut {NoStop}%
\bibitem [{\citenamefont {Esfarjani}\ \emph {et~al.}(2011)\citenamefont
  {Esfarjani}, \citenamefont {Chen},\ and\ \citenamefont
  {Stokes}}]{stokes_prb11}%
  \BibitemOpen
  \bibfield  {author} {\bibinfo {author} {\bibfnamefont {K.}~\bibnamefont
  {Esfarjani}}, \bibinfo {author} {\bibfnamefont {G.}~\bibnamefont {Chen}}, \
  and\ \bibinfo {author} {\bibfnamefont {H.~T.}\ \bibnamefont {Stokes}},\
  }\href {\doibase 10.1103/PhysRevB.84.085204} {\bibfield  {journal} {\bibinfo
  {journal} {Phys. Rev. B}\ }\textbf {\bibinfo {volume} {84}},\ \bibinfo
  {pages} {085204} (\bibinfo {year} {2011})}\BibitemShut {NoStop}%
\bibitem [{\citenamefont {Ganose}\ \emph {et~al.}(2021)\citenamefont {Ganose},
  \citenamefont {Park}, \citenamefont {Faghaninia}, \citenamefont
  {Woods-Robinson}, \citenamefont {Persson},\ and\ \citenamefont
  {Jain}}]{ganose_natcomm_21}%
  \BibitemOpen
  \bibfield  {author} {\bibinfo {author} {\bibfnamefont {A.~M.}\ \bibnamefont
  {Ganose}}, \bibinfo {author} {\bibfnamefont {J.}~\bibnamefont {Park}},
  \bibinfo {author} {\bibfnamefont {A.}~\bibnamefont {Faghaninia}}, \bibinfo
  {author} {\bibfnamefont {R.}~\bibnamefont {Woods-Robinson}}, \bibinfo
  {author} {\bibfnamefont {K.~A.}\ \bibnamefont {Persson}}, \ and\ \bibinfo
  {author} {\bibfnamefont {A.}~\bibnamefont {Jain}},\ }\href {\doibase
  10.1038/s41467-021-22440-5} {\bibfield  {journal} {\bibinfo  {journal}
  {Nature Communications}\ }\textbf {\bibinfo {volume} {12}},\ \bibinfo {pages}
  {2222} (\bibinfo {year} {2021})}\BibitemShut {NoStop}%
\bibitem [{\citenamefont {Zhou}\ \emph {et~al.}(2014)\citenamefont {Zhou},
  \citenamefont {Nielson}, \citenamefont {Xia},\ and\ \citenamefont
  {Ozoliņ{\v{s}}}}]{zhou_prl_14}%
  \BibitemOpen
  \bibfield  {author} {\bibinfo {author} {\bibfnamefont {F.}~\bibnamefont
  {Zhou}}, \bibinfo {author} {\bibfnamefont {W.}~\bibnamefont {Nielson}},
  \bibinfo {author} {\bibfnamefont {Y.}~\bibnamefont {Xia}}, \ and\ \bibinfo
  {author} {\bibfnamefont {V.}~\bibnamefont {Ozoliņ{\v{s}}}},\ }\href
  {\doibase 10.1103/PhysRevLett.113.185501} {\bibfield  {journal} {\bibinfo
  {journal} {Physical Review Letters}\ }\textbf {\bibinfo {volume} {113}},\
  \bibinfo {pages} {185501} (\bibinfo {year} {2014})}\BibitemShut {NoStop}%
\bibitem [{\citenamefont {Glassbrenner}\ and\ \citenamefont
  {Slack}(1964)}]{glassbrenner_64}%
  \BibitemOpen
  \bibfield  {author} {\bibinfo {author} {\bibfnamefont {C.~J.}\ \bibnamefont
  {Glassbrenner}}\ and\ \bibinfo {author} {\bibfnamefont {G.~A.}\ \bibnamefont
  {Slack}},\ }\href {\doibase 10.1103/PhysRev.134.A1058} {\bibfield  {journal}
  {\bibinfo  {journal} {Physical Review}\ }\textbf {\bibinfo {volume} {134}},\
  \bibinfo {pages} {A1058} (\bibinfo {year} {1964})}\BibitemShut {NoStop}%
\bibitem [{\citenamefont {Carlson}\ \emph {et~al.}(1965)\citenamefont
  {Carlson}, \citenamefont {Slack},\ and\ \citenamefont
  {Silverman}}]{carlson_65}%
  \BibitemOpen
  \bibfield  {author} {\bibinfo {author} {\bibfnamefont {R.~O.}\ \bibnamefont
  {Carlson}}, \bibinfo {author} {\bibfnamefont {G.~A.}\ \bibnamefont {Slack}},
  \ and\ \bibinfo {author} {\bibfnamefont {S.~J.}\ \bibnamefont {Silverman}},\
  }\href@noop {} {\bibfield  {journal} {\bibinfo  {journal} {J. Appl. Phys.}\
  }\textbf {\bibinfo {volume} {36}},\ \bibinfo {pages} {505} (\bibinfo {year}
  {1965})}\BibitemShut {NoStop}%
\bibitem [{\citenamefont {Morelli}\ \emph {et~al.}(2002)\citenamefont
  {Morelli}, \citenamefont {Heremans},\ and\ \citenamefont
  {Slack}}]{morelli_prb02}%
  \BibitemOpen
  \bibfield  {author} {\bibinfo {author} {\bibfnamefont {D.~T.}\ \bibnamefont
  {Morelli}}, \bibinfo {author} {\bibfnamefont {J.~P.}\ \bibnamefont
  {Heremans}}, \ and\ \bibinfo {author} {\bibfnamefont {G.~A.}\ \bibnamefont
  {Slack}},\ }\href {\doibase 10.1103/PhysRevB.66.195304} {\bibfield  {journal}
  {\bibinfo  {journal} {Phys. Rev. B}\ }\textbf {\bibinfo {volume} {66}},\
  \bibinfo {pages} {195304} (\bibinfo {year} {2002})}\BibitemShut {NoStop}%
\bibitem [{\citenamefont {Malhotra}\ and\ \citenamefont
  {Maldovan}(2016)}]{malhotra_scireps16}%
  \BibitemOpen
  \bibfield  {author} {\bibinfo {author} {\bibfnamefont {A.}~\bibnamefont
  {Malhotra}}\ and\ \bibinfo {author} {\bibfnamefont {M.}~\bibnamefont
  {Maldovan}},\ }\href {\doibase 10.1038/srep25818} {\bibfield  {journal}
  {\bibinfo  {journal} {Scientific Reports}\ }\textbf {\bibinfo {volume} {6}},\
  \bibinfo {pages} {25818} (\bibinfo {year} {2016})}\BibitemShut {NoStop}%
\bibitem [{\citenamefont {Sipahigil}\ \emph {et~al.}(2014)\citenamefont
  {Sipahigil}, \citenamefont {Jahnke}, \citenamefont {Rogers}, \citenamefont
  {Teraji}, \citenamefont {Isoya}, \citenamefont {Zibrov}, \citenamefont
  {Jelezko},\ and\ \citenamefont {Lukin}}]{sipahigil_prl_14}%
  \BibitemOpen
  \bibfield  {author} {\bibinfo {author} {\bibfnamefont {A.}~\bibnamefont
  {Sipahigil}}, \bibinfo {author} {\bibfnamefont {K.~D.}\ \bibnamefont
  {Jahnke}}, \bibinfo {author} {\bibfnamefont {L.~J.}\ \bibnamefont {Rogers}},
  \bibinfo {author} {\bibfnamefont {T.}~\bibnamefont {Teraji}}, \bibinfo
  {author} {\bibfnamefont {J.}~\bibnamefont {Isoya}}, \bibinfo {author}
  {\bibfnamefont {A.~S.}\ \bibnamefont {Zibrov}}, \bibinfo {author}
  {\bibfnamefont {F.}~\bibnamefont {Jelezko}}, \ and\ \bibinfo {author}
  {\bibfnamefont {M.~D.}\ \bibnamefont {Lukin}},\ }\href {\doibase
  10.1103/PhysRevLett.113.113602} {\bibfield  {journal} {\bibinfo  {journal}
  {Physical Review Letters}\ }\textbf {\bibinfo {volume} {113}},\ \bibinfo
  {pages} {113602} (\bibinfo {year} {2014})}\BibitemShut {NoStop}%
\bibitem [{\citenamefont {Sukachev}\ \emph {et~al.}(2017)\citenamefont
  {Sukachev}, \citenamefont {Sipahigil}, \citenamefont {Nguyen}, \citenamefont
  {Bhaskar}, \citenamefont {Evans}, \citenamefont {Jelezko},\ and\
  \citenamefont {Lukin}}]{sipahigil_prl_17}%
  \BibitemOpen
  \bibfield  {author} {\bibinfo {author} {\bibfnamefont {D.~D.}\ \bibnamefont
  {Sukachev}}, \bibinfo {author} {\bibfnamefont {A.}~\bibnamefont {Sipahigil}},
  \bibinfo {author} {\bibfnamefont {C.~T.}\ \bibnamefont {Nguyen}}, \bibinfo
  {author} {\bibfnamefont {M.~K.}\ \bibnamefont {Bhaskar}}, \bibinfo {author}
  {\bibfnamefont {R.~E.}\ \bibnamefont {Evans}}, \bibinfo {author}
  {\bibfnamefont {F.}~\bibnamefont {Jelezko}}, \ and\ \bibinfo {author}
  {\bibfnamefont {M.~D.}\ \bibnamefont {Lukin}},\ }\href {\doibase
  10.1103/PhysRevLett.119.223602} {\bibfield  {journal} {\bibinfo  {journal}
  {Physical Review Letters}\ }\textbf {\bibinfo {volume} {119}},\ \bibinfo
  {pages} {223602} (\bibinfo {year} {2017})}\BibitemShut {NoStop}%
\bibitem [{\citenamefont {Bluhm}\ \emph {et~al.}(2011)\citenamefont {Bluhm},
  \citenamefont {Foletti}, \citenamefont {Neder}, \citenamefont {Rudner},
  \citenamefont {Mahalu}, \citenamefont {Umansky},\ and\ \citenamefont
  {Yacoby}}]{bluhm_natphys_11}%
  \BibitemOpen
  \bibfield  {author} {\bibinfo {author} {\bibfnamefont {H.}~\bibnamefont
  {Bluhm}}, \bibinfo {author} {\bibfnamefont {S.}~\bibnamefont {Foletti}},
  \bibinfo {author} {\bibfnamefont {I.}~\bibnamefont {Neder}}, \bibinfo
  {author} {\bibfnamefont {M.}~\bibnamefont {Rudner}}, \bibinfo {author}
  {\bibfnamefont {D.}~\bibnamefont {Mahalu}}, \bibinfo {author} {\bibfnamefont
  {V.}~\bibnamefont {Umansky}}, \ and\ \bibinfo {author} {\bibfnamefont
  {A.}~\bibnamefont {Yacoby}},\ }\href {\doibase 10.1038/nphys1856} {\bibfield
  {journal} {\bibinfo  {journal} {Nature Physics}\ }\textbf {\bibinfo {volume}
  {7}},\ \bibinfo {pages} {109} (\bibinfo {year} {2011})}\BibitemShut {NoStop}%
\bibitem [{\citenamefont {Cao}\ \emph {et~al.}(2016)\citenamefont {Cao},
  \citenamefont {Li}, \citenamefont {Yu}, \citenamefont {Wang}, \citenamefont
  {Chen}, \citenamefont {Song}, \citenamefont {Xiao}, \citenamefont {Guo},
  \citenamefont {Jiang}, \citenamefont {Hu},\ and\ \citenamefont
  {Guo}}]{cao_prl_16}%
  \BibitemOpen
  \bibfield  {author} {\bibinfo {author} {\bibfnamefont {G.}~\bibnamefont
  {Cao}}, \bibinfo {author} {\bibfnamefont {H.~O.}\ \bibnamefont {Li}},
  \bibinfo {author} {\bibfnamefont {G.~D.}\ \bibnamefont {Yu}}, \bibinfo
  {author} {\bibfnamefont {B.~C.}\ \bibnamefont {Wang}}, \bibinfo {author}
  {\bibfnamefont {B.~B.}\ \bibnamefont {Chen}}, \bibinfo {author}
  {\bibfnamefont {X.~X.}\ \bibnamefont {Song}}, \bibinfo {author}
  {\bibfnamefont {M.}~\bibnamefont {Xiao}}, \bibinfo {author} {\bibfnamefont
  {G.~C.}\ \bibnamefont {Guo}}, \bibinfo {author} {\bibfnamefont {H.~W.}\
  \bibnamefont {Jiang}}, \bibinfo {author} {\bibfnamefont {X.}~\bibnamefont
  {Hu}}, \ and\ \bibinfo {author} {\bibfnamefont {G.~P.}\ \bibnamefont {Guo}},\
  }\href {\doibase 10.1103/PhysRevLett.116.086801} {\bibfield  {journal}
  {\bibinfo  {journal} {Physical Review Letters}\ }\textbf {\bibinfo {volume}
  {116}},\ \bibinfo {pages} {086801} (\bibinfo {year} {2016})}\BibitemShut
  {NoStop}%
\bibitem [{\citenamefont {Kresse}\ and\ \citenamefont
  {Hafner}(1993)}]{kresse93}%
  \BibitemOpen
  \bibfield  {author} {\bibinfo {author} {\bibfnamefont {G.}~\bibnamefont
  {Kresse}}\ and\ \bibinfo {author} {\bibfnamefont {J.}~\bibnamefont
  {Hafner}},\ }\href {\doibase 10.1103/PhysRevB.47.558} {\bibfield  {journal}
  {\bibinfo  {journal} {Phys. Rev. B}\ }\textbf {\bibinfo {volume} {47}},\
  \bibinfo {pages} {558} (\bibinfo {year} {1993})}\BibitemShut {NoStop}%
\bibitem [{\citenamefont {Kresse}\ and\ \citenamefont
  {Furthm{\"{u}}ller}(1996)}]{kresse96a}%
  \BibitemOpen
  \bibfield  {author} {\bibinfo {author} {\bibfnamefont {G.}~\bibnamefont
  {Kresse}}\ and\ \bibinfo {author} {\bibfnamefont {J.}~\bibnamefont
  {Furthm{\"{u}}ller}},\ }\href {\doibase
  https://doi.org/10.1016/0927-0256(96)00008-0} {\bibfield  {journal} {\bibinfo
   {journal} {Computational Materials Science}\ }\textbf {\bibinfo {volume}
  {6}},\ \bibinfo {pages} {15} (\bibinfo {year} {1996})}\BibitemShut {NoStop}%
\bibitem [{\citenamefont {Perdew}\ \emph {et~al.}(1996)\citenamefont {Perdew},
  \citenamefont {Burke},\ and\ \citenamefont {Ernzerhof}}]{PBE}%
  \BibitemOpen
  \bibfield  {author} {\bibinfo {author} {\bibfnamefont {J.~P.}\ \bibnamefont
  {Perdew}}, \bibinfo {author} {\bibfnamefont {K.}~\bibnamefont {Burke}}, \
  and\ \bibinfo {author} {\bibfnamefont {M.}~\bibnamefont {Ernzerhof}},\ }\href
  {\doibase 10.1103/PhysRevLett.77.3865} {\bibfield  {journal} {\bibinfo
  {journal} {Physical Review Letters.}\ }\textbf {\bibinfo {volume} {77}},\
  \bibinfo {pages} {3865} (\bibinfo {year} {1996})}\BibitemShut {NoStop}%
\bibitem [{\citenamefont {Baroni}\ \emph {et~al.}(2001)\citenamefont {Baroni},
  \citenamefont {de~Gironcoli}, \citenamefont {Dal~Corso},\ and\ \citenamefont
  {Giannozzi}}]{baroni01}%
  \BibitemOpen
  \bibfield  {author} {\bibinfo {author} {\bibfnamefont {S.}~\bibnamefont
  {Baroni}}, \bibinfo {author} {\bibfnamefont {S.}~\bibnamefont
  {de~Gironcoli}}, \bibinfo {author} {\bibfnamefont {A.}~\bibnamefont
  {Dal~Corso}}, \ and\ \bibinfo {author} {\bibfnamefont {P.}~\bibnamefont
  {Giannozzi}},\ }\href {https://link.aps.org/doi/10.1103/RevModPhys.73.515}
  {\bibfield  {journal} {\bibinfo  {journal} {RMP}\ }\textbf {\bibinfo {volume}
  {73}},\ \bibinfo {pages} {515} (\bibinfo {year} {2001})}\BibitemShut
  {NoStop}%
\bibitem [{\citenamefont {Togo}\ and\ \citenamefont {Tanaka}(2015)}]{togo15}%
  \BibitemOpen
  \bibfield  {author} {\bibinfo {author} {\bibfnamefont {A.}~\bibnamefont
  {Togo}}\ and\ \bibinfo {author} {\bibfnamefont {I.}~\bibnamefont {Tanaka}},\
  }\href {\doibase 10.1016/j.scriptamat.2015.07.021} {\bibfield  {journal}
  {\bibinfo  {journal} {Scripta Materialia}\ }\textbf {\bibinfo {volume}
  {108}},\ \bibinfo {pages} {1} (\bibinfo {year} {2015})}\BibitemShut {NoStop}%
\bibitem [{\citenamefont {Jain}\ \emph {et~al.}(2013)\citenamefont {Jain},
  \citenamefont {Ong}, \citenamefont {Hautier}, \citenamefont {Chen},
  \citenamefont {Richards}, \citenamefont {Dacek}, \citenamefont {Cholia},
  \citenamefont {Gunter}, \citenamefont {Skinner}, \citenamefont {Ceder},\ and\
  \citenamefont {Persson}}]{materialsproject}%
  \BibitemOpen
  \bibfield  {author} {\bibinfo {author} {\bibfnamefont {A.}~\bibnamefont
  {Jain}}, \bibinfo {author} {\bibfnamefont {S.~P.}\ \bibnamefont {Ong}},
  \bibinfo {author} {\bibfnamefont {G.}~\bibnamefont {Hautier}}, \bibinfo
  {author} {\bibfnamefont {W.}~\bibnamefont {Chen}}, \bibinfo {author}
  {\bibfnamefont {W.~D.}\ \bibnamefont {Richards}}, \bibinfo {author}
  {\bibfnamefont {S.}~\bibnamefont {Dacek}}, \bibinfo {author} {\bibfnamefont
  {S.}~\bibnamefont {Cholia}}, \bibinfo {author} {\bibfnamefont
  {D.}~\bibnamefont {Gunter}}, \bibinfo {author} {\bibfnamefont
  {D.}~\bibnamefont {Skinner}}, \bibinfo {author} {\bibfnamefont
  {G.}~\bibnamefont {Ceder}}, \ and\ \bibinfo {author} {\bibfnamefont {K.~A.}\
  \bibnamefont {Persson}},\ }\href {\doibase 10.1063/1.4812323} {\bibfield
  {journal} {\bibinfo  {journal} {APL Materials}\ }\textbf {\bibinfo {volume}
  {1}},\ \bibinfo {pages} {011002} (\bibinfo {year} {2013})}\BibitemShut
  {NoStop}%
\bibitem [{\citenamefont {Chiarotti}\ and\ \citenamefont
  {Goletti}(2005)}]{encyclopedia}%
  \BibitemOpen
  \bibfield  {author} {\bibinfo {author} {\bibfnamefont {G.}~\bibnamefont
  {Chiarotti}}\ and\ \bibinfo {author} {\bibfnamefont {C.}~\bibnamefont
  {Goletti}},\ }in\ \href {\doibase 10.1016/B0-12-369401-9/00476-9} {\emph
  {\bibinfo {booktitle} {Encyclopedia of Condensed Matter Physics}}}\ (\bibinfo
   {publisher} {Elsevier},\ \bibinfo {year} {2005})\ pp.\ \bibinfo {pages}
  {133--144}\BibitemShut {NoStop}%
\bibitem [{\citenamefont {Maradudin}\ and\ \citenamefont
  {Fein}(1962)}]{maradudin62}%
  \BibitemOpen
  \bibfield  {author} {\bibinfo {author} {\bibfnamefont {A.~A.}\ \bibnamefont
  {Maradudin}}\ and\ \bibinfo {author} {\bibfnamefont {A.~E.}\ \bibnamefont
  {Fein}},\ }\href {\doibase 10.1103/PhysRev.128.2589} {\bibfield  {journal}
  {\bibinfo  {journal} {Phys. Rev.}\ }\textbf {\bibinfo {volume} {128}},\
  \bibinfo {pages} {2589} (\bibinfo {year} {1962})}\BibitemShut {NoStop}%
\bibitem [{\citenamefont {Canonico}\ \emph {et~al.}(2002)\citenamefont
  {Canonico}, \citenamefont {Poweleit}, \citenamefont {Men{\'{e}}ndez},
  \citenamefont {Debernardi}, \citenamefont {Johnson},\ and\ \citenamefont
  {Zhang}}]{canonico_prl_02}%
  \BibitemOpen
  \bibfield  {author} {\bibinfo {author} {\bibfnamefont {M.}~\bibnamefont
  {Canonico}}, \bibinfo {author} {\bibfnamefont {C.}~\bibnamefont {Poweleit}},
  \bibinfo {author} {\bibfnamefont {J.}~\bibnamefont {Men{\'{e}}ndez}},
  \bibinfo {author} {\bibfnamefont {A.}~\bibnamefont {Debernardi}}, \bibinfo
  {author} {\bibfnamefont {S.~R.}\ \bibnamefont {Johnson}}, \ and\ \bibinfo
  {author} {\bibfnamefont {Y.-H.}\ \bibnamefont {Zhang}},\ }\href {\doibase
  10.1103/PhysRevLett.88.215502} {\bibfield  {journal} {\bibinfo  {journal}
  {Physical Review Letters}\ }\textbf {\bibinfo {volume} {88}},\ \bibinfo
  {pages} {215502} (\bibinfo {year} {2002})}\BibitemShut {NoStop}%
\bibitem [{\citenamefont {Lindsay}(2016)}]{lindsay_prb16}%
  \BibitemOpen
  \bibfield  {author} {\bibinfo {author} {\bibfnamefont {L.}~\bibnamefont
  {Lindsay}},\ }\href {\doibase 10.1103/PhysRevB.94.174304} {\bibfield
  {journal} {\bibinfo  {journal} {Phys. Rev. B}\ }\textbf {\bibinfo {volume}
  {94}},\ \bibinfo {pages} {174304} (\bibinfo {year} {2016})}\BibitemShut
  {NoStop}%
\bibitem [{\citenamefont {Veldhorst}\ \emph {et~al.}(2014)\citenamefont
  {Veldhorst}, \citenamefont {Hwang}, \citenamefont {Yang}, \citenamefont
  {Leenstra}, \citenamefont {de~Ronde}, \citenamefont {Dehollain},
  \citenamefont {Muhonen}, \citenamefont {Hudson}, \citenamefont {Itoh},
  \citenamefont {Morello},\ and\ \citenamefont {Dzurak}}]{veldhorst_natnano14}%
  \BibitemOpen
  \bibfield  {author} {\bibinfo {author} {\bibfnamefont {M.}~\bibnamefont
  {Veldhorst}}, \bibinfo {author} {\bibfnamefont {J.~C.~C.}\ \bibnamefont
  {Hwang}}, \bibinfo {author} {\bibfnamefont {C.~H.}\ \bibnamefont {Yang}},
  \bibinfo {author} {\bibfnamefont {A.~W.}\ \bibnamefont {Leenstra}}, \bibinfo
  {author} {\bibfnamefont {B.}~\bibnamefont {de~Ronde}}, \bibinfo {author}
  {\bibfnamefont {J.~P.}\ \bibnamefont {Dehollain}}, \bibinfo {author}
  {\bibfnamefont {J.~T.}\ \bibnamefont {Muhonen}}, \bibinfo {author}
  {\bibfnamefont {F.~E.}\ \bibnamefont {Hudson}}, \bibinfo {author}
  {\bibfnamefont {K.~M.}\ \bibnamefont {Itoh}}, \bibinfo {author}
  {\bibfnamefont {A.}~\bibnamefont {Morello}}, \ and\ \bibinfo {author}
  {\bibfnamefont {A.~S.}\ \bibnamefont {Dzurak}},\ }\href {\doibase
  10.1038/nnano.2014.216} {\bibfield  {journal} {\bibinfo  {journal} {Nature
  Nanotechnology}\ }\textbf {\bibinfo {volume} {9}},\ \bibinfo {pages} {981}
  (\bibinfo {year} {2014})}\BibitemShut {NoStop}%
\bibitem [{\citenamefont {Inyushkin}\ \emph {et~al.}(2003)\citenamefont
  {Inyushkin}, \citenamefont {Taldenkov}, \citenamefont {Yakubovsky},
  \citenamefont {Markov}, \citenamefont {Moreno-Garsia},\ and\ \citenamefont
  {Sharonov}}]{gaas_enrichment}%
  \BibitemOpen
  \bibfield  {author} {\bibinfo {author} {\bibfnamefont {A.~V.}\ \bibnamefont
  {Inyushkin}}, \bibinfo {author} {\bibfnamefont {A.~N.}\ \bibnamefont
  {Taldenkov}}, \bibinfo {author} {\bibfnamefont {A.~Y.}\ \bibnamefont
  {Yakubovsky}}, \bibinfo {author} {\bibfnamefont {A.~V.}\ \bibnamefont
  {Markov}}, \bibinfo {author} {\bibfnamefont {L.}~\bibnamefont
  {Moreno-Garsia}}, \ and\ \bibinfo {author} {\bibfnamefont {B.~N.}\
  \bibnamefont {Sharonov}},\ }\href {\doibase 10.1088/0268-1242/18/7/315}
  {\bibfield  {journal} {\bibinfo  {journal} {Semiconductor Science and
  Technology}\ }\textbf {\bibinfo {volume} {18}},\ \bibinfo {pages} {685}
  (\bibinfo {year} {2003})}\BibitemShut {NoStop}%
\bibitem [{\citenamefont {Alaie}\ \emph {et~al.}(2015)\citenamefont {Alaie},
  \citenamefont {Goettler}, \citenamefont {Su}, \citenamefont {Leseman},
  \citenamefont {Reinke},\ and\ \citenamefont {El-Kady}}]{alaie_natcomm15}%
  \BibitemOpen
  \bibfield  {author} {\bibinfo {author} {\bibfnamefont {S.}~\bibnamefont
  {Alaie}}, \bibinfo {author} {\bibfnamefont {D.~F.}\ \bibnamefont {Goettler}},
  \bibinfo {author} {\bibfnamefont {M.}~\bibnamefont {Su}}, \bibinfo {author}
  {\bibfnamefont {Z.~C.}\ \bibnamefont {Leseman}}, \bibinfo {author}
  {\bibfnamefont {C.~M.}\ \bibnamefont {Reinke}}, \ and\ \bibinfo {author}
  {\bibfnamefont {I.}~\bibnamefont {El-Kady}},\ }\href {\doibase
  10.1038/ncomms8228} {\bibfield  {journal} {\bibinfo  {journal} {Nature
  Communications}\ }\textbf {\bibinfo {volume} {6}},\ \bibinfo {pages} {7228}
  (\bibinfo {year} {2015})}\BibitemShut {NoStop}%
\bibitem [{\citenamefont {Teichert}\ \emph {et~al.}(1995)\citenamefont
  {Teichert}, \citenamefont {MacKay}, \citenamefont {Savage}, \citenamefont
  {Lagally}, \citenamefont {Brohl},\ and\ \citenamefont
  {Wagner}}]{teichert_apl95}%
  \BibitemOpen
  \bibfield  {author} {\bibinfo {author} {\bibfnamefont {C.}~\bibnamefont
  {Teichert}}, \bibinfo {author} {\bibfnamefont {J.~F.}\ \bibnamefont
  {MacKay}}, \bibinfo {author} {\bibfnamefont {D.~E.}\ \bibnamefont {Savage}},
  \bibinfo {author} {\bibfnamefont {M.~G.}\ \bibnamefont {Lagally}}, \bibinfo
  {author} {\bibfnamefont {M.}~\bibnamefont {Brohl}}, \ and\ \bibinfo {author}
  {\bibfnamefont {P.}~\bibnamefont {Wagner}},\ }\href {\doibase
  10.1063/1.113978} {\bibfield  {journal} {\bibinfo  {journal} {Applied Physics
  Letters}\ }\textbf {\bibinfo {volume} {66}},\ \bibinfo {pages} {2346}
  (\bibinfo {year} {1995})}\BibitemShut {NoStop}%
\bibitem [{\citenamefont {Swartz}\ and\ \citenamefont
  {Pohl}(1989)}]{swartz_pohl_revmodphys_89}%
  \BibitemOpen
  \bibfield  {author} {\bibinfo {author} {\bibfnamefont {E.~T.}\ \bibnamefont
  {Swartz}}\ and\ \bibinfo {author} {\bibfnamefont {R.~O.}\ \bibnamefont
  {Pohl}},\ }\href {\doibase 10.1103/RevModPhys.61.605} {\bibfield  {journal}
  {\bibinfo  {journal} {Reviews of Modern Physics}\ }\textbf {\bibinfo {volume}
  {61}},\ \bibinfo {pages} {605} (\bibinfo {year} {1989})}\BibitemShut
  {NoStop}%
\bibitem [{\citenamefont {Harrelson}\ \emph {et~al.}()\citenamefont
  {Harrelson}, \citenamefont {Inzani},\ and\ \citenamefont
  {Griffin}}]{harrelson_interface}%
  \BibitemOpen
  \bibfield  {author} {\bibinfo {author} {\bibfnamefont {T.~F.}\ \bibnamefont
  {Harrelson}}, \bibinfo {author} {\bibfnamefont {K.}~\bibnamefont {Inzani}}, \
  and\ \bibinfo {author} {\bibfnamefont {S.~M.}\ \bibnamefont {Griffin}},\
  }\href@noop {} {\bibinfo  {journal} {Submitted}\ }\BibitemShut {NoStop}%
\bibitem [{\citenamefont {Debernardi}\ \emph {et~al.}(1995)\citenamefont
  {Debernardi}, \citenamefont {Baroni},\ and\ \citenamefont
  {Molinari}}]{debernardi_prl95}%
  \BibitemOpen
\bibfield  {journal} {  }\bibfield  {author} {\bibinfo {author} {\bibfnamefont
  {A.}~\bibnamefont {Debernardi}}, \bibinfo {author} {\bibfnamefont
  {S.}~\bibnamefont {Baroni}}, \ and\ \bibinfo {author} {\bibfnamefont
  {E.}~\bibnamefont {Molinari}},\ }\href {\doibase 10.1103/PhysRevLett.75.1819}
  {\bibfield  {journal} {\bibinfo  {journal} {Physical Review Letters}\
  }\textbf {\bibinfo {volume} {75}},\ \bibinfo {pages} {1819} (\bibinfo {year}
  {1995})}\BibitemShut {NoStop}%
\bibitem [{\citenamefont {Glensk}\ \emph {et~al.}(2019)\citenamefont {Glensk},
  \citenamefont {Grabowski}, \citenamefont {Hickel}, \citenamefont
  {Neugebauer}, \citenamefont {Neuhaus}, \citenamefont {Hradil}, \citenamefont
  {Petry},\ and\ \citenamefont {Leitner}}]{glensk_prl19}%
  \BibitemOpen
  \bibfield  {author} {\bibinfo {author} {\bibfnamefont {A.}~\bibnamefont
  {Glensk}}, \bibinfo {author} {\bibfnamefont {B.}~\bibnamefont {Grabowski}},
  \bibinfo {author} {\bibfnamefont {T.}~\bibnamefont {Hickel}}, \bibinfo
  {author} {\bibfnamefont {J.}~\bibnamefont {Neugebauer}}, \bibinfo {author}
  {\bibfnamefont {J.}~\bibnamefont {Neuhaus}}, \bibinfo {author} {\bibfnamefont
  {K.}~\bibnamefont {Hradil}}, \bibinfo {author} {\bibfnamefont
  {W.}~\bibnamefont {Petry}}, \ and\ \bibinfo {author} {\bibfnamefont
  {M.}~\bibnamefont {Leitner}},\ }\href {\doibase
  10.1103/PhysRevLett.123.235501} {\bibfield  {journal} {\bibinfo  {journal}
  {Physical Review Letters}\ }\textbf {\bibinfo {volume} {123}},\ \bibinfo
  {pages} {235501} (\bibinfo {year} {2019})}\BibitemShut {NoStop}%
\bibitem [{\citenamefont {Gourley}\ and\ \citenamefont
  {Drummond}(1986)}]{gourley86}%
  \BibitemOpen
  \bibfield  {author} {\bibinfo {author} {\bibfnamefont {P.~L.}\ \bibnamefont
  {Gourley}}\ and\ \bibinfo {author} {\bibfnamefont {T.~J.}\ \bibnamefont
  {Drummond}},\ }\href {\doibase 10.1063/1.97126} {\bibfield  {journal}
  {\bibinfo  {journal} {Applied Physics Letters}\ }\textbf {\bibinfo {volume}
  {49}},\ \bibinfo {pages} {489} (\bibinfo {year} {1986})}\BibitemShut
  {NoStop}%
\end{thebibliography}%
\makeatletter\@input{xx.tex}\makeatother
\end{document}


\title{Supplementary Information for First principles investigation of scattering processes in phonon-based quantum sensors}
\author{Thomas F. Harrelson}
\affiliation{Materials Science Division, Lawrence Berkeley National Laboratory, Berkeley, CA 94720, USA}
\affiliation{Molecular Foundry, Lawrence Berkeley National Laboratory, Berkeley, CA 94720, USA}
\author{Ibrahim Hajar}
\affiliation{Molecular Foundry, Lawrence Berkeley National Laboratory, Berkeley, CA 94720, USA}
\affiliation{Department of Mathematics, University of California - San Diego, CA 92093}
\affiliation{The Blackett Laboratory, Imperial College London, Prince Consort Road, London SW7 2BW, U.K.}
\author{Omar A. Ashour}
\affiliation{Materials Science Division, Lawrence Berkeley National Laboratory, Berkeley, CA 94720, USA}
\affiliation{Molecular Foundry, Lawrence Berkeley National Laboratory, Berkeley, CA 94720, USA}
\affiliation{Department of Physics, University of California, Berkeley, California 94720, USA}
\author{Sinéad M. Griffin}
\affiliation{Materials Science Division, Lawrence Berkeley National Laboratory, Berkeley, CA 94720, USA}
\affiliation{Molecular Foundry, Lawrence Berkeley National Laboratory, Berkeley, CA 94720, USA}

\maketitle

\section{Phonon bandstructures} \label{si:sec:pbands}

\begin{figure}[h!]
    \centering
    \includegraphics[width=\linewidth]{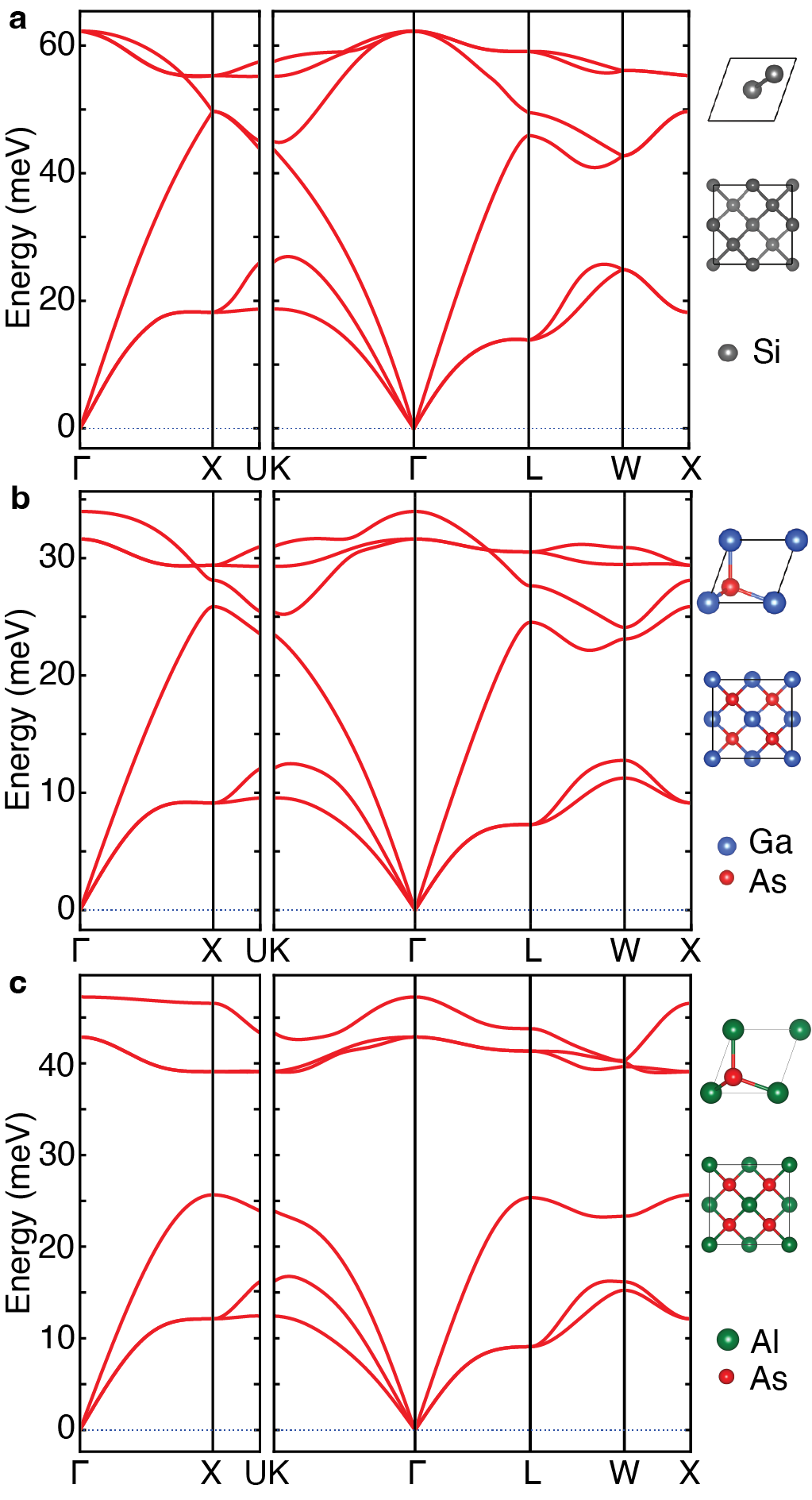}
    \caption{Phonon band structures for (a) Si, (b) GaAs, and (c) AlAs. The primitive unit cell for each crystal is shown on the top right of the respective band structure diagram, and the conventional unit cell underneath. The lattice constants are 5.47~$\text{\AA}$, 5.76~\AA, and 5.73~\AA, respectively.}
    \label{fig:crys_bands}
\end{figure}

We first use DFT calculations to optimize the structural parameters of silicon, gallium arsenide, and aluminum arsenide crystals with the convergence parameters described in the main text. All three materials adopt cubic crystal structures with Si having a diamond (Fd$\bar{3}$m) structure and GaAs/AlAs having a zincblende structure (F$\bar{4}$3m).  Our calculated phonon bandstructures for the three crystals are plotted in Figure~\ref{fig:crys_bands}, consistent with previously reported phonon bandstructures. We observe that the acoustic modes for each material are similar, due to the similarities between the crystal structures. The main difference between the phonon bandstructures is in the optical branches because GaAs and AlAs are polar materials (Si is nonpolar with one type of atom in the unit cell), which generates a splitting between the longitudinal and transverse optical modes at  $\Gamma$. This is easily seen in Figure~\ref{fig:crys_bands} along the W-X symmetry line, in which the optical and acoustic branches are degenerate for Si, but split in GaAs and AlAs. 


\section{Anharmonic Lifetime Calculation}

Anharmonic scattering rates were calculated by constructing a $2\times2\times2$ supercell of the conventional standard cell, and approximating the second and third order force constants via the finite difference method as implemented within Phono3py. Having calculated the third-order force constants, the phonon imaginary self-energy was calculated via the formalism in Refs.~\cite{maradudin62, phono3py}, as implemented within Phono3py. The formula for the imaginary self-energy is,
\begin{multline}
    \Gamma_{\lambda_0}(\omega)= \frac{18\pi}{\hbar^2}\sum_{\lambda_1, \lambda_2} \left|\Phi_{\lambda_0,\lambda_1,\lambda_2}\right|^2 \{(n_{\lambda_0}+n_{\lambda_1}+1) \delta(\omega - \omega_{\lambda_1}-\omega_{\lambda_2}) \\ + (n_{\lambda_1}-n_{\lambda_2})\left[\delta(\omega + \omega_{\lambda_1} - \omega_{\lambda_2}) - \delta(\omega - \omega_{\lambda_1} + \omega_{\lambda_2})\right]\}
\end{multline}
in which $\tau$ is the phonon lifetime, $\Phi_{\lambda_0,\lambda_1,\lambda_2}$ is the three phonon matrix element, $\lambda$ is the phonon index that combines both the mode index and k-point for brevity, $n_{\lambda_0}$ is the average occupation number for the phonon defined by $\lambda_0$, and $\omega_{\lambda_0}$ is the frequency for the phonon defined by $\lambda_0$.

\section{Isotopic Lifetime Analysis} \label{si:sec:isotopes}

\begin{figure}
    \centering
    \includegraphics[width=\linewidth]{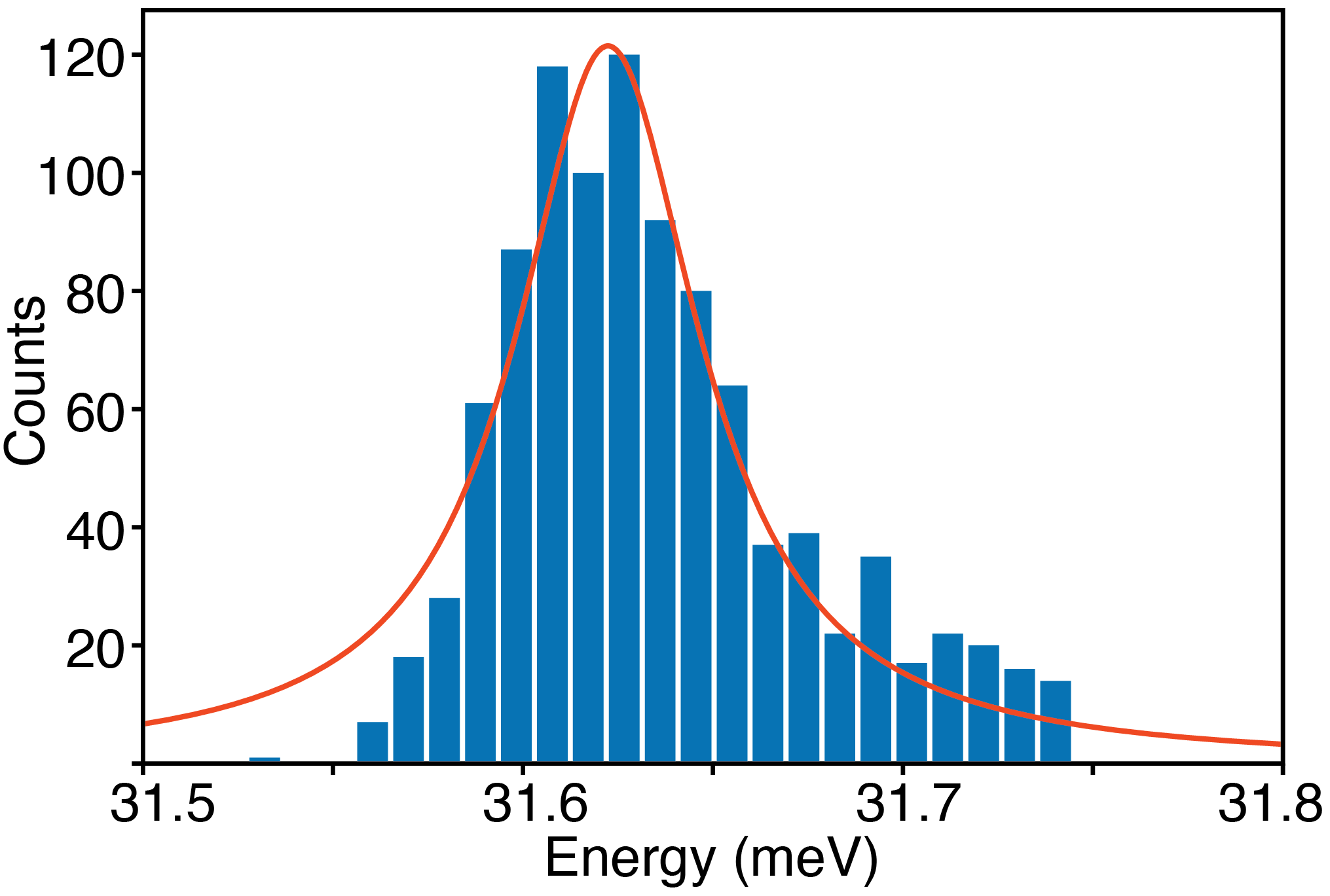}
    \caption{Energetic variation of a gamma-point optical phonon mode for GaAs due to random isotopic impurities.}
    \label{fig:isotope-hist}
\end{figure}

To determine the importance of this scattering mechanism, we analyze the percentage of scattering events that do not conserve energy. To do this, we calculate phonon frequencies while varying the isotopic composition. We started with a GaAs $2\times2\times2$ supercell of the conventional unit cell (64 atoms), and randomly assigned isotopic masses to each Gallium atom in the cell while maintaining the natural distribution of isotopes (92.2\% $^{28}$Si, 4.7\% $^{29}$Si, 3.1\% $^{30}$Si, and 60.1\% $^{69}$Ga, 39.9\% $^{71}$Ga). We next calculated the frequencies for each random isotopic configuration. The phonon eigenvectors for a given configuration are multiplied with the eigenvectors for the structure with uniform masses to create a set of scores, which generates a map between the isotopically perturbed phonons and the uniform sample. Using this map, we have a set of isotopically perturbed frequencies for each phonon branch and k-point in the Brillouin zone. We plot a histogram of the gamma-point phonons for an optical branch of GaAs in Figure~\ref{fig:isotope-hist}.

We observe that the optical phonon frequency is 31.6~meV, corresponding to a lifetime of 70.0~ps. The isotopic lifetime calculated within Phono3py is 79.7~ps, which indicates that nearly all of the isotopic scattering for this $\Gamma$-point optical mode is explained by the energetic broadening. The numerically calculated isotopic scattering rate from Figure~\ref{fig:isotope-hist} is greater than the scattering rate calculated from Phono3py; we expect that by including more random samples in the histogram, the distribution would get smoother and converge to the Phono3py value. 

To better understand this result, we consider the isotopic scattering rate given in Ref.~\cite{Tamura1983IsotopeGe},
\begin{equation} \label{si:eq:iso_scattering_rate}
    \tau^{-1}_\lambda(\omega) = \frac{2\pi}{N}\omega_\lambda^2 \sum_{\lambda', i} g_{2,i} \left|\Vec{\epsilon}^*_{i,\lambda'}\cdot \Vec{\epsilon}_{i,\lambda}\right|^2\delta(\omega-\omega_{\lambda'})
\end{equation}
where $\omega_\lambda$ is the frequency of mode $\lambda$, $\epsilon$ is the phonon eigenvector, $g_{2,i}$ is the isotopic mass variance of atom $i$, $N$ is the number of unit cells in the crystal, and $\lambda$ represents the combined phonon index (including the wave-vector and branch index). In this equation, we see that the original mode $\lambda$ is uniformly coupled to all other $\lambda'$ modes. The only selection rules are driven by the inner product between phonon eigenvectors, which indicates that coupling only exists between modes of the same type. For example, a longitudinal phonon can couple to other longitudinal modes, and likewise for a transverse mode to other transverse modes. Therefore, when phonons are scattered by isotopic variation, the initial phonon can transfer energy to other modes with different energies, showing that thermalization solely due to isotopic scattering is possible given enough time. 

This result may appear to be at odds with the isotopic scattering rate formula derived from Fermi's golden rule, equivalent to evaluating Eq.~\eqref{si:eq:iso_scattering_rate} at $\omega_\lambda$, which enforces energy conservation. In fact, the isotopic scattering rate used in Phono3py for thermal transport calculations is $\tau_\lambda^{-1}(\omega_\lambda)$ from Eq.~\eqref{si:eq:iso_scattering_rate}. It is tempting to interpret Fermi's golden rule as proof that the energy is conserved through the scattering interaction. However, this interpretation is only correct to zeroth order, or for systems with discrete energy levels. However, phonon states in extended media live on a continuous manifold of quantum numbers (k-points in the Brillouin zone). For example, the interaction between two degenerate phonon states causes a splitting between the states, but these new perturbed states become degenerate with different states in the Brillouin zone with which the new states can couple to, causing an additional splitting, and the process repeats itself. Thus, the perturbation to the Hamiltonian potentially causes coupling between all states in the Brillouin zone. Therefore, we should not expect that a state initially at some energy, $\omega_o$, will stay at that energy as the system evolves in time. However, the decay rate from the \textit{initial} state is perfectly well-described by Fermi's golden rule; the rule does not include how the energies of the \textit{final} states are perturbed by the scattering interaction. Another way of stating this is that the final phonon states in the unperturbed picture are not eigenstates of the perturbed Hamiltonian, so the final states are a linear combination of eigenstates that need not have the same energy. The expectation value of the energy for that linear combination of eigenstates has the same energy as the final state in the unperturbed picture.

\section{Kirchhoff approach to surface scattering} \label{si:sec:kirchhoff}

The contribution of diffusive scattering due to surfaces with a finite roughness has been described by Ref.~\cite{malhotra_scireps16},
\begin{equation}
    s = e^{-4\eta^2k^2\cos^2\theta}
\end{equation}
where $s$ is the probability of specular reflection of phonon upon interacting with an interface, $\eta$ is the surface roughness parameter (with units of length), $k$ is the phonon wavevector, and $\theta$ is the angle between the $k$ and a vector perpendicular to the surface. Since we are interested in the average surface scattering rates, we average over all $\theta$ directions, which yields,
\begin{equation}
    \langle s \rangle_\theta = \int_0^{\pi/2} e^{-4\eta^2k^2\cos^2\theta} \sin\theta \text{d}\theta
\end{equation}
After substituting $u=\cos\theta$, we obtain,
\begin{equation}
    \langle s \rangle_\theta = \int_0^{1} e^{-4\eta^2k^2u^2} \text{d}u
\end{equation}
substituting again with $\Tilde{u} = 2\eta ku$ yields,
\begin{equation} \label{si:eq:kirchhoff-subd}
    \langle s \rangle_\theta = \frac{1}{2\eta k} \int_0^{2\eta k} e^{-\Tilde{u}^2} \text{d}\Tilde{u}
\end{equation}
By using the error function,
\begin{equation}
    \text{erf} (x) = \frac{2}{\sqrt{\pi}} \int_0^x e^{-y^2}\text{d}y
\end{equation}
we can rewrite Eq.~\eqref{si:eq:kirchhoff-subd} as,
\begin{equation}
    \langle s \rangle_\theta = \frac{\sqrt{\pi}}{4\eta k} \text{erf} (2\eta k)
\end{equation}
which is used in the main text to determine the rate of non-specular scattering due to surface interactions.

\section{Thermal flux of phonons in diffusive regime} \label{si:sec:diffusive}

\begin{figure}
    \centering
    \includegraphics{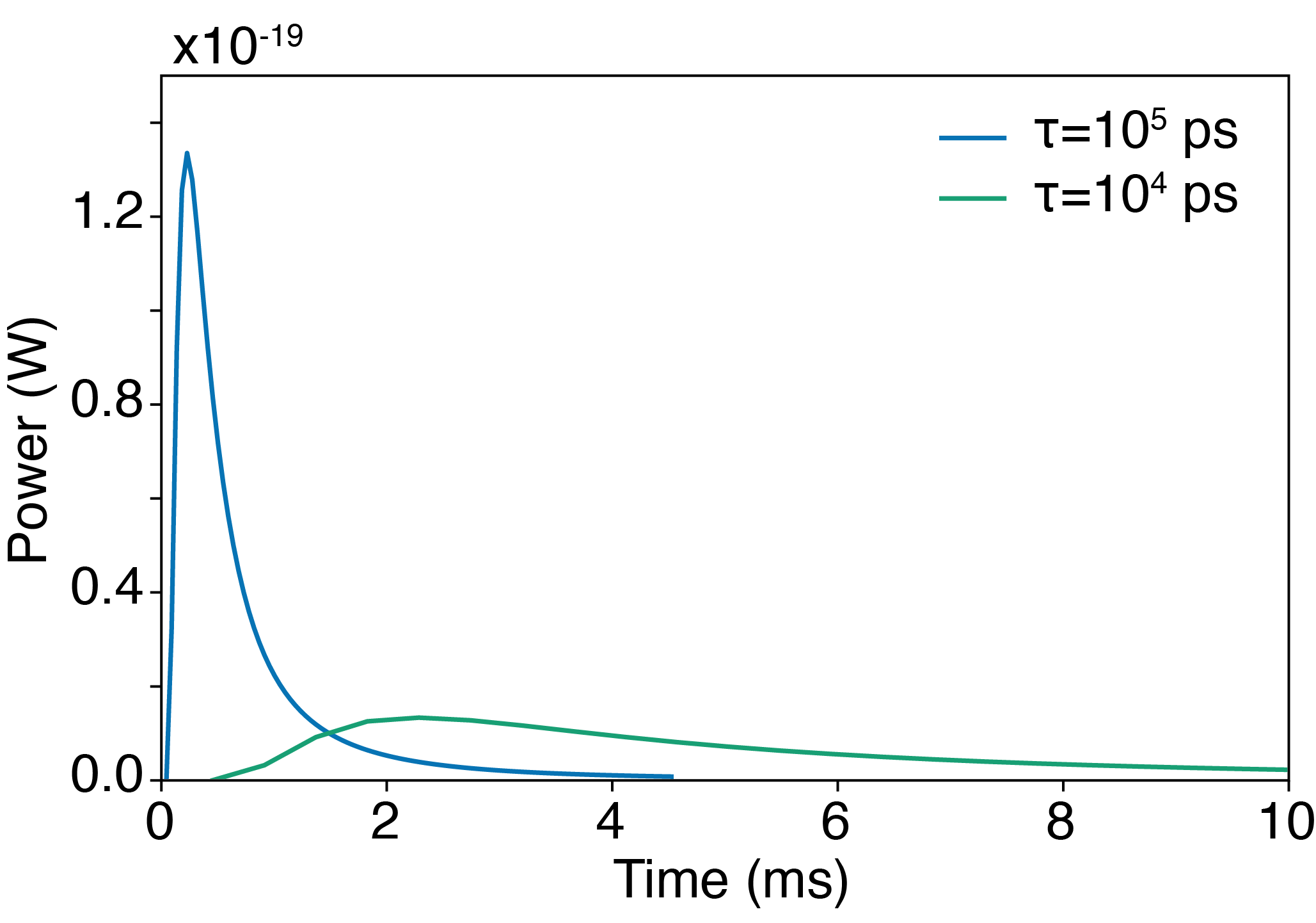}
    \caption{Diffusive thermal power as a function of time for 10~meV phonon in a 50~cm$^3$ absorber crystal, with varying relaxation times.}
    \label{si:fig:diff_trans}
\end{figure}

In the limit of no phonon scattering, the phonon eigenstates are completely extended Bloch waves of atomic displacement amplitudes. Introducing scattering has the effect of localizing these states in real space. Assuming an initial distribution function of a delta function in real space, a phonon distribution function evolves diffusively in time according to,
\begin{equation} \label{si:eq:diff_transport}
    n(r,t) = \left(\frac{1}{4\pi D t}\right)^{3/2} \exp{\frac{-r^2}{4Dt}}
\end{equation}
in which $n(r,t)$ is the distribution of phonons (similar to a number density) at position $r$ and time $t$, and $D$ is the diffusion coefficient of the phonon distribution. At time $t=0$, we recover the initial delta function. We are implicitly setting the axes' origin to the center of the initial distribution. It is an approximation to start with an infinitely narrow phonon distribution, but this becomes a good approximation in the limit in which the length scales we are interested in are much larger than the localization of the initial distribution. In this case, the two characteristic length scales are the crystal size, $L$, and the localization length of the initial phonon distribution, $\ell=\sqrt{\langle r^2 \rangle}$. Thus the condition in which Eq.~\eqref{si:eq:diff_transport} is valid is,
\begin{equation}
    \frac{L}{\ell} >> 1
\end{equation}
We expect that $\ell \sim v_g \tau$, where typical values of $v_g$ are $\sim1$~km/s, providing a relationship between the relaxation time $\tau$ and $\ell$. If $L\sim 1$~cm, then for $\ell\sim 0.01$~cm, then $\tau\sim 10^{-7}$~s, or $\tau\sim 10^5$~ps. From Figure 2 in the main text, we see that there are many phonon modes that satisfy this condition, meaning Eq.~\eqref{si:eq:diff_transport} is valid for those phonon modes. Using a constitutive diffusive relation (Fick's law), we connect the diffusive current to the gradient of the phonon number density,
\begin{equation}
    \vec{j}_n(r,t) = -D \vec{\nabla} n(r,t)
\end{equation}
where $\vec{j}_n(r,t)$ is the time-dependent diffusive phonon current. Given $n$ from Eq.~\eqref{si:eq:diff_transport}, we find that,
\begin{equation} \label{si:eq:diff_current}
    \vec{j}_n(r,t) = -\left(\frac{1}{4\pi D t}\right)^{3/2} \left(\frac{\vec{r}}{2t}\right)\exp{\frac{-r^2}{4Dt}}
\end{equation}
Evaluating $\vec{j}_n(r=L, t)$, we find that at long times, $j_n\propto t^{-5/2}$, and $\lim_{t\rightarrow 0} j_n(L,t) = 0$, so that that there is a finite lifetime to the diffusive flux signal at the boundary of the absorber material. We show the thermal power as a function of time for $\tau=10^5$~ps and $\tau=10^4$~ps in Figure~\ref{si:fig:diff_trans}. 

In this Figure~\ref{si:fig:diff_trans}, we find that the power signal reaches $\sim10^{-19}$~W for $\sim1$~ms when the phonon relaxation time is 10$^5$~ps. Given the noise-equivalent power is $1.5\times10^{-18}$~W/Hz$^{1/2}$, we find that the effective noise power for 1~kHz bandwidth is $\sim 5\times10^{-17}$~W, which is more than two orders of magnitude greater than the predicted signal power. When the phonon relaxation time is $10^{4}$~ps, then the signal is $\sim 10^{-20}$~W, but survives for $\sim 10$~ms. Using a bandwidth of 100~Hz, the effective noise power is $1.5\times10^{-17}$~W, which, again, is more than two orders of magnitude greater than the predicted signal power. Thus, in the diffusive limit of transport, the signal cannot be detected with the state of current readout technologies. We add that the readout surface coverage of 2.7\% is very inefficient because diffusive transport does not benefit from reduced surface coverage. Increasing surface coverage to 100\% increases the signal power by a factor of 37, but even then the signal power remains significantly smaller than the noise power. We note that this analysis assumes that the diffusive phonon population does not down-convert to lower energies. Given that the time it takes the signal to reach the readout interface is $\sim$1~ms, we expect that nearly all phonons will anharmonically down-convert to lower energies, further reducing the signal intensity.

\bibliography{ref2}
\makeatletter\@input{yy.tex}\makeatother